\documentclass[]{interact}
\usepackage{url} 
\usepackage{epstopdf}% To incorporate .eps illustrations using PDFLaTeX, etc.
\usepackage[caption=false]{subfig}% Support for small, `sub' figures and tables
\usepackage{float}
\usepackage{enumitem}

\usepackage{xcolor}
\usepackage{soul}
\definecolor{r1}{RGB}{0,255,255}% Reviewer 1
\definecolor{r2}{RGB}{255,165,0}% Reviewer 2

\definecolor{revblue}{RGB}{0,90,200}% Revised/new text in this revision round
\usepackage[longnamesfirst,sort]{natbib}% Citation support using natbib.sty
\bibpunct[, ]{(}{)}{;}{a}{,}{,}% Citation support using natbib.sty
\renewcommand\bibfont{\fontsize{10}{12}\selectfont}% To set the list of references in 10 point font using natbib.sty

\theoremstyle{plain}% Theorem-like structures provided by amsthm.sty

\theoremstyle{definition}

\theoremstyle{remark}

\begin{document}

% \articletype{}% Specify the article type or omit as appropriate

\title{Harmonic Eigenspace: A Web-based Application for Navigating and Composing Microtonal Harmony}

% \author{
% \name{A Author\textsuperscript{a}\thanks{Contact A Author. Email: author@university.org } and B Author\textsuperscript{b}}
% \affil{\textsuperscript{a}, \textsuperscript{b} }
% }

\author{
\name{David Dalmazzo\textsuperscript{a}\thanks{Contact: Dr. Dalmazzo. Email: david.cabrera@upf.edu} and Ken D\'eguernel\textsuperscript{b}}
\affil{\textsuperscript{a}Music Technology Group, Pompeu Fabra University, Barcelona, Spain}
\affil{\textsuperscript{b}Univ. Lille, CNRS, Centrale Lille, UMR 9189 CRIStAL, F-59000 Lille, France}
}

\maketitle

\begin{abstract}
This paper presents a web-based application for navigating and composing microtonal harmony, built on the \emph{Harmonic Eigenspace}, a four-dimensional psychoacoustically grounded space in which tetrad chord types are located by their spectral dissonance profiles, computed with Sethares's roughness/dissonance model. 
The coordinate system is transposition-invariant:
a coordinate triple $(\alpha, \beta, \gamma)$ locates the three upper notes in relation to the root, so a chord quality corresponds to a direction in the space, the invariant ray along which transposition acts, while the root frequency sets the scale.
The dissonance field over these coordinates can be computed at any register;
the locations of its local minima are register-invariant, as they arise from partial-coincidence ratio conditions.
The dissonance volume contains 100 local minima that align with just-intonation intervals and act as landmarks, organising the space into basins around the most consonant tetrads.
We embed tetrads from three tonal equal temperaments as discrete lattices within this continuous volume. 
The application presents this space through two components: 
the \emph{Harmonic Eigenspace} as a navigable 3D visualisation of the dissonance volume in which all nodes are playable, and a \emph{Modal Studio} that extends modal interchange logic to the ten-gradation interval vocabulary of 53-TET.
The application also functions as a MIDI controller with MIDI Polyphonic Expression support, which can be integrated into any digital audio workstation that supports this format.
A listening study with 31 participants used both scenes of the application: listeners were first exposed to isolated 53-TET chords, recording how these new qualities are perceived alongside chords familiar from Western practice, such as the maj7 and the m7; they then rated chord progressions composed in the  \textit{Modal Studio}, measuring their acceptance/rejection of microtonal chord progressions encountered for the first time.
\end{abstract}


% Plan:
% 1) Introduction and SOTA. ✅

% 2) Theoretical Concept. ✅
% Mathematical Foundation ✅
% We start from 2D to 3D to 4D Eigenspace. Explain the math of the vector space. ✅
% We have to formulate the Eigen principle, ✅

% 3) The design of the application  ✅

% 4) Musical harmonic extension principle ✅
% TODO: add examples on a web and explain the new harmon ic options

% 5) Audition/Listening test: ✅
% Perception test to support the idea of dissonant vs consonant musical perception of 53-TET possibilities. 

% 6) Discussion and Conclusion ✅

% Introduction section — ANIMA Harmonic Eigenspace paper
% ------------------------------------------------------------------------------
\section{Introduction}
\label{sec:introduction}
% 1) Microtonality
% 2) This paper presents a web-app application. 
% 3) Eigenspace. Perception object.
% 4) Modal Studio.
% 5) Listening test.

Approached from the Western perspective, microtonality opens new creative possibilities with a richer harmonic vocabulary that expands the standard 12-tone equal temperament (12-TET), which offers two third-quality gradations, minor and major. In 31-TET we find five: sub-minor, minor, neutral, major, and super-major; and in 53-TET, ten: sub-minor, down-minor, minor, up-minor, minor-neutral, major-neutral, down-major, major, up-major, and super-major. These creative possibilities open questions about how to explore and navigate the trajectories of chord relations. From this starting point, we systematically explore the vocabulary and propose an instrument in which the new qualities can be located, compared, and heard.

Thus, this paper presents a web application that offers two scenes for exploring microtonal harmonies from two complementary perspectives. The first scene is a four-dimensional space formed by visualising and sonifying a cloud of dissonant/consonant points. Those points are located in a proposed space we call the \textit{Harmonic Eigenspace}, formed by the degree of sensory dissonance, computed from the roughness between partials, from a chosen root (for example, A3 = 220~Hz) across the ratio span of one octave above it. The second scene is a grid we call the \textit{Modal Studio}, where a chord progression can be formed using modes derived from the 31- and 53-tone equal temperaments, and modal interchange substitutions are generated and presented in the application. The method's finer scale resolution lets the 12-TET modes be deformed into new formations, introducing new intervallic relations and chord qualities. We conduct a listening test to expose participants to both scenes and observe how listeners perceive basic examples in terms of novelty or musical tolerance/acceptance of the proposed chords and harmonies.

The Harmonic Eigenspace principle rests on a psychoacoustic basis for the perception of consonance and dissonance, arising from the pattern of roughness produced by the simultaneous partials of the sounding tones. This perceptive dissonance relationship was measured and formalised by \citet{plomp1965tonal} and extended to complex timbres by \citet{sethares2005tuning}. Hence, we construct a space in which each tetrad chord type occupies a position determined by its psychoacoustic dissonance profile: a root and three upper voices span four frequency dimensions, and the three interval ratios above the root form the coordinates, so chord types populate a three-dimensional volume within a four-dimensional pitch space. In closed position within one octave, this structure is the seventh chord, in which the qualities of the third, the fifth, and the seventh act simultaneously. Therefore, a chord quality corresponds to an invariant direction of pitch space under transposition; the local minima of the dissonance field serve as landmarks that define distances to the chord types.

From here, we locate chords in 12-, 31- and 53-TET. 53-TET's close approximation to both the just perfect fifth and the major third makes it consistent with both Pythagorean and five-limit just intonation \citep{tanaka1890studien, fokker1963refinement}. The Harmonic Eigenspace (hereafter, the Eigenspace) provides a psychoacoustically rather than music-theoretically grounded coordinate system for them: chord types are placed as discrete objects without borrowing principles from a different tuning system.

However, it is important to clarify that a chord quality is a perceptual object. When a listener hears a major seventh chord as stable or a diminished chord as tense, that perception is shaped by cultural context, as well as by the timbral characteristics of the musical instrument. The perceived dissonance varies across musical traditions and practices. For instance, in Colombian \textit{Cumbia}, dominant seventh chords can function as a stable tonic rather than a tension requiring a resolution; the rhythm plays that role of tension and release \citep{ochoa2016cumbia, valencia2019libro}. In the blues, major-minor harmonic dualities do not conform to the harmonic principles of Western tonal practice, yet listeners perceive their chord progressions as coherent and expressive \citep{mulholland2013berklee, martin2023tonic}. Many musical traditions worldwide do not organise pitch simultaneously into chord qualities at all, relying instead on melodic, rhythmic, or timbral dimensions of consonance and tension. This is why we also propose a second scene where user can explore and listen to 31- and 53-TET chord progressions in a more idiomatic way called the \textit{Modal Studio}.

In the \textit{Modal Studio}, we approach microtonality from the perspective of Western modal-interchange principles applied in 12-TET. The harmonic navigation is proposed by forming new modes and therefore new tetrads: the different gradations of intervals from 31- and 53-TET can be heard and tested using a \textit{Scale Editor} interface that forms new dispositions of scales, mirroring the twelve notes used in Western modal practice while offering a controlled surface on which to experiment and navigate new harmonic trajectories. The standard modes can also be replicated, but there is a rich collection of tetrad types that deviate from 12-TET, the 152 catalogued in Section~\ref{chord_quality_extensions}; many of those chords occupy regions of the Eigenspace where no 12-TET chord exists. These generated progressions open new paths for musicians exploring microtonality.

% The interactive application built around the Eigenspace assigns the computer a specific role: assisting a composer in navigating the harmonic space by locating chord types according to their perceptual properties, auditioning them, and constructing progressions through modal interchange logic extended to the interval vocabulary of 53-TET. This aligns with the ANIMA project's commitment to tools that support rather than replace human creative agency \citep{dalmazzo2025computer}.

The research questions are:

\begin{enumerate}[label=\alph*)]

% \item What is the topological structure of the four-dimensional dissonance volume computed from the Sethares roughness model, and how do 12-TET, 31-TET, and 53-TET tetrad types distribute as discrete lattices within its continuous landscape?

\item Can chord quality be defined as a position in a psychoacoustically grounded space, such that the tetrad vocabularies of 12-, 31-, and 53-TET, and the continuum beyond them, can be located, compared, and audibly navigated in one coordinate system?

\item When modal interchange logic is extended to the ten-gradation interval vocabulary of 53-TET, do the resulting chord progressions sustain perceived harmonic coherence and musicality, and how do these perceptual ratings vary with the tonic chord's proximity to familiar 12-TET qualities?

\end{enumerate}

The contribution is twofold. The technical contribution is the instrument: a web application that renders the dissonance volume as a navigable, sounding visualisation, extends modal interchange to the 31- and 53-TET vocabulary in its \textit{Modal Studio}, and integrates with standard production environments through MIDI/MPE. The theoretical contribution is the foundation the instrument is built on, the Eigenspace: a transposition-invariant coordinate space of tetrad types, grounded in the dissonance function of \citet{sethares2005tuning}, in which 12-, 31-, and 53-TET coexist as discrete lattices and can be compared without presupposing any scale theory. The listening study documents how listeners, with their own cultural formation, respond to chords and progressions drawn from the space.

The paper is organised as follows. Section~\ref{sec:literature_review} surveys related work on geometric models of harmony, interactive harmonic exploration systems, and computational approaches to microtonality. Section~\ref{sec:mathematical_foundations} develops the mathematical foundations of the Eigenspace, from the pairwise dissonance model for sinusoids through to the four-dimensional dissonance surface for tetrads. Section~\ref{sec:harmonic_extension}  presents the harmonic principle proposed. The design and implementation of the interactive application are described in Section~\ref{sec:application}. Section~\ref{sec:listening_test} presents the listening test we performed. Section~\ref{sec:conclusion} discusses implications and outlines directions for future work.

The application is available at \url{https://dazzid.github.io/ANIMA_Harmonic_Eigenspace/}.

% ------------------------------------------------------------------------------
\section{Literature Review}
\label{sec:literature_review}

% \subsection{Psychoacoustic Foundations of Consonance and Dissonance}

% The perceptual distinction between consonance and dissonance has a quantitative foundation established by \citet{plomp1965tonal}, who measured listener judgements of roughness for pairs of pure sinusoids and showed that perceived dissonance depends primarily on the frequency separation between tones relative to the auditory critical bandwidth. Dissonance rises from zero at unison, reaches a maximum near one quarter of the critical bandwidth, and then decays as the two tones become individually resolvable.

% \citet{sethares2005tuning} extended this model from pure sinusoids to complex tones by summing the pairwise dissonances across all interacting partials of two simultaneously sounding notes. The central consequence is that the locations of dissonance minima depend directly on the spectral content of the timbre. For harmonic timbres, these minima align closely with the simple integer ratios of just intonation, recovering the familiar intervals of Western tonal music. For inharmonic timbres, the minima shift accordingly. This establishes that consonance is not a fixed property of interval ratios but emerges from the relationship between tuning and timbre, and motivates the use of this model as a basis for evaluating the perceptual status of microtonal intervals without circular reference to existing scale theory.

We first survey work on the computational modelling of spatial representations of harmony, then interactive systems for exploring harmony, and finally, microtonality applied to musicking as a cultural practice.

\subsection{Geometric and Spatial Models of Harmony in 12-TET}

\citet{balzano1980group} framed harmonic structures using cyclic group theory, showing that properties of the diatonic scale and the system of major and minor keys can be derived from the algebraic structure of 12-fold pitch-class space, and argued that the same reasoning extends to other equal temperaments. \citet{chew2007out} proposed the Spiral Array model, placing pitches, chords, and keys on a helical surface in three-dimensional space, with higher-level structures represented as convex combinations of their constituent lower-level elements. This continuous representation reduces key-finding and chord identification to nearest-neighbour searches in the spiral geometry, and has been validated against empirical data on tonal pitch perception \citep{lerdahl2001tonal}.

A closely related tradition represents harmonic relationships through the Tonnetz, a two-dimensional lattice in which pitch classes are nodes, triads occupy triangular cells, and Neo-Riemannian operations correspond to reflections across shared edges. \citet{cohn1997neo} formalised this structure and showed that voice leading between triads maps directly onto geometric proximity within it. \citet{tymoczko2012generalized} extended the framework to a broader class of chords and tuning systems, while \citet{milne2016empirically} tested several models of perceived triadic distance against listener ratings, finding that Tonnetz distance and spectral pitch-class distance were near-equivalent predictors and nearly perfectly correlated with each other, while voice-leading models performed poorly, suggesting that the Tonnetz derives its predictive power from psychoacoustic rather than contrapuntal structure. The present work differs in its object: rather than testing models of the perceived distance between pairs of 12-TET triads, it constructs a coordinate space of tetrad types in which 12-, 31-, and 53-TET can be located and compared. Topological approaches have extended spatial thinking to chord progressions. \citet{bigo2015topological} represented progressions as paths through simplicial complexes, where vertices correspond to pitch classes and higher-dimensional simplices correspond to chords, making voice-leading structure geometrically explicit. These spatial models share an important limitation for the present purpose: their geometry is constructed from music-theoretic axioms (fifth-relatedness, scale membership, voice-leading parsimony) rather than from a perceptual measurement.

Recent empirical work supports grounding harmonic geometry in perceptual measurement: \citet{frank2026modeling} show that spectral interference is the primary axis along which listeners vary in their pleasantness judgements of chords, accounting for the largest proportion of systematic between-listener variation, an effect that strengthens with musical expertise.

\subsection{Interactive Systems for Harmonic Exploration}

\citet{holland1994learning} introduced Harmony Space, an early interactive system that uses spatial metaphors derived from \citet{balzano1980group} and \citet{longuet1962letter} theories to make harmonic relationships navigable. Chords and keys are arranged on a two-dimensional grid, and users explore tonal structure through direct manipulation, making the system effective for both novices and experienced musicians. More recent systems have incorporated data-driven generation. \citet{huang2016chordripple} built \textit{ChordRipple}, which recommends chord continuations by learning statistical patterns from a corpus, extending a novice composer's harmonic vocabulary beyond familiar progressions. Transformer-based approaches \citep{choi2016text, wu2020jazz} have since pushed sequential chord generation further, though these systems remain bound to 12-TET and provide no perceptual account of why particular chords sound consonant or tense.

% The application described in this paper is distinguished from these systems by the generative principle of its harmonic space. Rather than recommending statistically likely continuations, it maps the perceptual dissonance landscape of 53-TET as a navigable four-dimensional space, allowing a composer to locate chords by their psychoacoustic properties and trace paths that constitute harmonic progressions. The modal interchange logic described in \citet{dalmazzo2025computer} provides a theory-grounded mechanism for chord substitution within this landscape.

\subsection{Cultural Practices of Microtonal Music}

Microtonal practice has a long and continuous history in both Western and non-Western music. Harry Partch spent decades building instruments capable of realising his 43-tone scale, grounding an entire compositional language in just intonation ratios that standard Western tuning cannot express \citep{gilmore2003climate}. In Europe, Manfred Stahnke and his Hamburg school have pursued microtonal composition since the 1980s, working with just intonation and 31-tone frameworks. Outside the Western tradition, 53 equal divisions of the octave serve as the reference unit of modern Turkish makam theory, embracing its historical Pythagorean tone systems with less than a cent of error \citep{yarman2007comparative}.

Contemporary microtonal practice is growing and increasingly visible. Jacob Collier has brought microtonal harmony into mainstream popular music, introducing intermediate steps within standard chord progressions and adjusting voicing to align consonant sonorities outside the 12-TET ratio. Online communities have accelerated this dissemination; for example, the educators and composers of \textit{Hear Between The Lines}\footnote{\url{https://www.youtube.com/@HearBetweenTheLines} (last accessed July 2026)} offer tutorials and compositions using 31-TET custom instruments. Zheanna Erose\footnote{\url{https://www.youtube.com/@ZheannaErose} (last accessed July 2026)} uses microtonal keyboards and explores non-Western modes. Mike Battaglia\footnote{\url{https://www.youtube.com/@MikeBattagliaMusic} (last accessed July 2026)} blends Western harmonic concepts and complex voicing in blues, jazz, and his own compositions. The composer, instrumentalist, and singer Maddie Ashman\footnote{\url{https://maddieashman.bandcamp.com} (last accessed July 2026)} has grown in popularity, using a custom-tuned keyboard, a guitar with a slider, and a cello. The Quebec-based microtonal rock band \textit{Angine de Poitrine}\footnote{\url{https://anginedepoitrine.bandcamp.com/album/vol-ii} (last accessed July 2026)}, using guitar, bass, electronics (pedals) in 24-TET, and drums; has reached 3.5M views on the YouTube channel \textit{KEXP}\footnote{\url{https://www.youtube.com/channel/UC3I2GFN_F8WudD_2jUZbojA} (last accessed July 2026)}. The \textit{xenharmonic} movement\footnote{\url{https://en.wikipedia.org/wiki/Xenharmonic_music} (last accessed July 2026)}, with contributions from artists including Elaine Walker, Birdhouse, and Stephen James Taylor\footnote{Elaine Walker, ``What Is Xenharmonic Music?'', NewMusicBox, New Music USA, 2017, \url{https://newmusicusa.org/nmbx/what-is-xenharmonic-music/} (last accessed July 2026)}, using microtonal exploration for electronic music. 

Instrument development has followed: microtunable synthesisers date back at least to the Yamaha DX7-II (1986), whose user-tuning tables made non-12-TET scales widely available; the \textit{Lumatone}\footnote{\url{https://www.lumatone.io} (last accessed July 2026)} isomorphic keyboard offers a hardware interface capable of mapping 31-TET and 53-TET tuning systems, though its cost remains a barrier to wider adoption. The \textit{Roli}\footnote{\url{https://roli.com} (last accessed July 2026)} keyboard introduced the possibility of bending notes on a 12-TET-based keyboard by enabling an interactive surface to express intermediate steps between keys. The same company introduced the MPE\footnote{\url{https://roli.com/mpe} (last accessed July 2026)} formatting. The piano keyboard called \textit{Osmose}\footnote{\url{https://www.expressivee.com/2-osmose} (last accessed July 2026)} allows pitch bend expressive control on keys using the MPE format.

\subsection{Microtonality in Computation and Practice}

The theoretical grounding for 53-TET goes back to \citet{holder1731treatise} and was formalised by \citet{fokker1955equal, fokker1963refinement}, who developed the antan\'{a}iresis approach to identifying equal temperaments that closely approximate just intonation intervals. \citet{muzzulini2020isaac} documents Newton's investigation of equal divisions of the octave from 15 to 612 parts, with particular attention to the 53-division for its accuracy in approximating the perfect fifth and the major third. 53-TET tempers out both the schisma ($32805{:}32768 \approx 2.0$ cents, the residue by which eight just fifths and a just major third exceed five octaves) and the kleisma ($15625{:}15552 \approx 8.1$ cents, the residue between six just minor thirds and an octave plus a fifth), making it simultaneously consistent with Pythagorean and five-limit just intonation \citep{tanaka1890studien}.

In computational music, \citet{anders2010computational} developed a constraint-based system for microtonal composition supporting multiple temperaments, though it was discontinued with its host platform. \citet{hirai2023microtonal} provides a dataset of microtonal MIDI and audio recordings to support machine learning work in this domain. The present application addresses a gap none of these fill: a real-time interactive tool grounded in psychoacoustic dissonance in which the perceptual landscape itself determines the structure of the navigable space, applied specifically to the harmonic possibilities of 53-TET.

% ------------------------------------------------------------------------------
\section{Mathematical Foundations}
\label{sec:mathematical_foundations}
This section develops the mathematical basis of the application's harmonic map, the Harmonic Eigenspace, from the ground up. We begin applying the Sethares formula foundation, explained in sections \ref{subsec:sensory_dissonance_sinusoids} and \ref{subsec:dissonance_curves}: two pure sinusoids sounding together, and the perceptual roughness that arises between them. From there, we apply the model to a third simultaneous frequency, forming a 3D topology with consonance valleys and dissonance mountains (section \ref{subsec:triad_surface}). Then, we add the fourth note, forming a four-dimensional dissonance surface that forms the coordinate system of the Harmonic Eigenspace (section \ref{subsec:Eigenspace}). The underlying principle throughout is that dissonance is the sum of pairwise roughness across all interacting partials.

\subsection{Sensory dissonance for pairs of sinusoids}
\label{subsec:sensory_dissonance_sinusoids}

The mathematical foundation of the Harmonic Eigenspace begins with a simple perceptual fact: when two pure sinusoids are sounded simultaneously, listeners perceive roughness that depends on how far apart the two frequencies are. \citet{plomp1965tonal} measured this systematically and found a consistent shape: dissonance is zero at unison, rises to a maximum when the two tones are separated by roughly one quarter of the auditory critical bandwidth, and then falls back toward zero as they become individually resolvable by the cochlea. \citet{sethares2005tuning} fitted a difference of two exponentials to these curves:

\begin{equation}
    d(x) = e^{-b_1 x} - e^{-b_2 x}
    \label{eq:dissonance_basic}
\end{equation}

\noindent where $x = |f_2 - f_1|$ is the frequency difference between the two sinusoids. The two exponentials pull in opposite directions: the faster-decaying term ($b_2 = 5.75$) drives the curve up from zero, while the slower-decaying term ($b_1 = 3.5$) pulls it back down. Their difference produces a single peak, whose location is the point of maximum dissonance:

\begin{equation}
    x^{*} = \frac{\ln(b_1 / b_2)}{b_1 - b_2} \approx 0.22
    \label{eq:xstar}
\end{equation}

\noindent in normalised units. The values of $b_1$ and $b_2$ were obtained by least-squares fit to the Plomp--Levelt model.

This is sufficient for a single fixed frequency, but the auditory critical bandwidth is not constant: it grows with increasing frequency. A separation of 20~Hz causes strong roughness around 100~Hz but is barely perceptible around 4~kHz. To account for this, the frequency axis must be rescaled locally depending on where on the spectrum the interaction occurs. Sethares approximates the critical bandwidth with a linear function of frequency, $s_1 f_1 + s_2$, which gives the scaling factor:

\begin{equation}
    s = \frac{x^{*}}{s_1 f_1 + s_2}
    \label{eq:scaling}
\end{equation}

\noindent with $s_1 = 0.021$ and $s_2 = 19$~Hz. The constant $s_2$ sets the baseline bandwidth at very low frequencies; $s_1$ captures its linear growth with frequency. Both are again fitted to experimental data. The constant $x^{*}$ of Equation~\ref{eq:xstar} enters the numerator of Equation~\ref{eq:scaling} so that, after rescaling, the point of maximum dissonance falls at the same fraction of the critical bandwidth at every register. Substituting $s$ into the dissonance function, and weighting by the loudness of the quieter of the two sinusoids (since a very soft component contributes little roughness regardless of its frequency proximity), gives the full pairwise dissonance for a pair of sinusoids:

\begin{equation}
    d(f_1, f_2, a_1, a_2) = a_{12} \left[ e^{-b_1\, s\, (f_2 - f_1)} - e^{-b_2\, s\, (f_2 - f_1)} \right]
    \label{eq:dissonance_full}
\end{equation}

\noindent where $f_1 < f_2$ and $a_{12} = \min(a_1, a_2)$. Equation~\ref{eq:dissonance_full} is stated for two sinusoids; Section~\ref{subsec:dissonance_curves} applies it pairwise to the sinusoidal partials of complex tones.

The perceptual reality of this roughness measure has a biological correlate: \citet{shamma2019spectro} demonstrate that pitch and beating percepts for both resolved and unresolved harmonics are unified early in the auditory pathway through spectro-temporal templates, whose dendritic filtering directly constrains the saliency and frequency range of the roughness sensation that the Sethares model quantifies.

% ------------------------------------------------------------------------------
\subsection{Dissonance curves for complex tones}
\label{subsec:dissonance_curves}

In practice, most musical sounds contain multiple partials. We model a tone $F$ as a set of $n$ steady sinusoidal partials at frequencies $f_1 < f_2 < \cdots < f_n$ with loudnesses $a_1, a_2, \ldots, a_n$; the model is thus a steady-state description, assigning a single dissonance value to a stationary spectrum, and temporal properties of sound (attack, decay, duration) lie outside its scope. The intrinsic dissonance of $F$ is the sum of the pairwise dissonances across all its own partials:

\begin{equation}
    D_F = \frac{1}{2} \sum_{i=1}^{n} \sum_{j=1}^{n} d(f_i,\, f_j,\, a_i,\, a_j)
    \label{eq:intrinsic_dissonance}
\end{equation}

When two notes with the same spectrum $F$ are sounded simultaneously at a frequency ratio $\alpha$, the transposed spectrum $\alpha F$ contains partials at $\alpha f_1, \alpha f_2, \ldots, \alpha f_n$. The total dissonance as a function of $\alpha$ is:

\begin{equation}
    D_F(\alpha) = D_F + D_{\alpha F} + \sum_{i=1}^{n} \sum_{j=1}^{n} d(f_i,\, \alpha f_j,\, a_i,\, a_j)
    \label{eq:dissonance_curve}
\end{equation}

\noindent where $D_F$ and $D_{\alpha F}$ are the intrinsic dissonances of each note, and the double sum captures the interactions between partials of the two notes.

\begin{figure}[t]
    \centering
    \includegraphics[width=0.9\columnwidth]{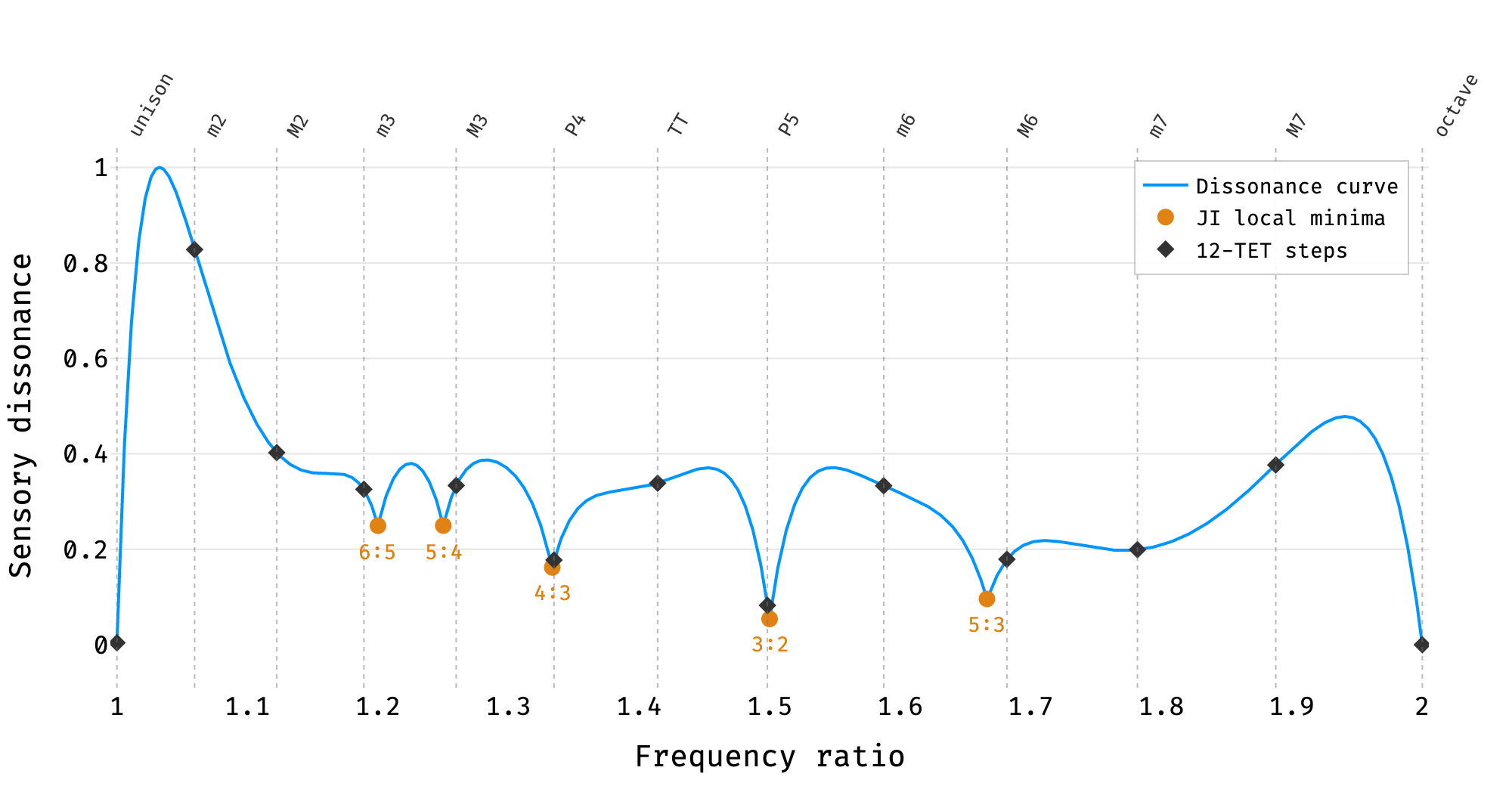}
    \caption{Sensory dissonance $D_F(\alpha)$ over one octave for a harmonic timbre with six partials at integer multiples of a 500~Hz fundamental and loudnesses $0.88^{k}$, the configuration of \citet{sethares2005tuning}; the curve is normalised so that the unison is 0 and the maximum is 1. Orange markers: local minima, located at the just-intonation ratios $6\!:\!5$, $5\!:\!4$, $4\!:\!3$, $3\!:\!2$, and $5\!:\!3$ (partial coincidences); no minimum arises at $8\!:\!5$ or $7\!:\!4$, which would require the eighth and seventh partial. Black diamonds: the 12-TET steps, with dissonance values unison $0.00$, m2 $0.83$, M2 $0.40$, m3 $0.33$, M3 $0.33$, P4 $0.18$, tritone $0.34$, P5 $0.08$, m6 $0.33$, M6 $0.18$, m7 $0.20$, M7 $0.38$, octave $0.00$. The perfect fourth and fifth fall close on their just minima; the thirds and the major sixth lie on the shoulders slightly off theirs; the minor second, the tritone, and the major seventh lie in regions of similar dissonance.}
    
    \label{fig:1d_dissonance}
\end{figure}

Figure~\ref{fig:1d_dissonance} shows $D_F(\alpha)$ across one octave, $\alpha \in [1,\, 2]$, for a harmonic timbre with six partials at integer multiples of a 500~Hz fundamental and loudnesses $0.88^{k}$, $k = 0, \ldots, 5$, the configuration used by \citet{sethares2005tuning} for his reference dissonance curve. The curve exhibits local minima at frequency ratios of just intonation: the minor third ($6\!:\!5 = 1.200$), major third ($5\!:\!4 = 1.250$), perfect fourth ($4\!:\!3 \approx 1.333$), perfect fifth ($3\!:\!2 = 1.500$), and major sixth ($5\!:\!3 \approx 1.667$), in addition to the unison ($1\!:\!1$) and octave ($2\!:\!1$). These minima arise because, at simple integer ratios, the upper partials of the two tones coincide or fall far apart, producing little beating. When the ratio deviates from these values, nearby partials fall within each other's critical bands and generate roughness. Minima can only arise at ratios where partials of the two tones coincide: with six partials there is no coincidence at the minor sixth ($8\!:\!5$) or the septimal minor seventh ($7\!:\!4$), which would require the eighth and the seventh partial, respectively. For the same reason, the number of minima is finite: Sethares bounds it by $2n^{2}$ for a spectrum of $n$ partials. Partials of zero or negligible loudness contribute nothing, since each pairwise term in Equation~\ref{eq:dissonance_full} is weighted by the loudness of the quieter partial.

This result establishes a direct link between the spectral content of a sound and the intervals that will be perceived as consonant. For harmonic timbres, the minima recover the familiar intervals of Western tonal music. For inharmonic timbres, the minima shift to entirely different locations, implying that consonance is not a fixed property of interval ratios but a function of the relationship between tuning and timbre. The construction is therefore not assumption-free: it inherits the roughness model of \citet{sethares2005tuning}, its fitted psychoacoustic parameters, and the choice of a harmonic timbre, and the alignment of the minima with just intonation throughout this paper is a consequence of these choices; Section~\ref{sec:conclusion} discusses the inharmonic case.

\subsection{The dissonance surface for triads}
\label{subsec:triad_surface}

The dissonance model extends naturally to chords. For three simultaneous notes at intervals $1$, $r$, and $s$, the total dissonance is:

\begin{equation}
    D_F(r, s) = D_F(r) + D_F(s) + D_{rF}\!\left(\frac{s}{r}\right)
    \label{eq:chord_dissonance}
\end{equation}

\noindent where $D_F(r)$ is the dissonance at interval $r$, $D_F(s)$ at interval $s$, and $D_{rF}(s/r)$ captures the interaction between the second and third notes. Generalisations to $m$ simultaneous tones follow the same principle of summing the dissonances across all pairs of sounding partials. 

\begin{figure}[H]
    \centering
    \includegraphics[width=0.7\columnwidth]{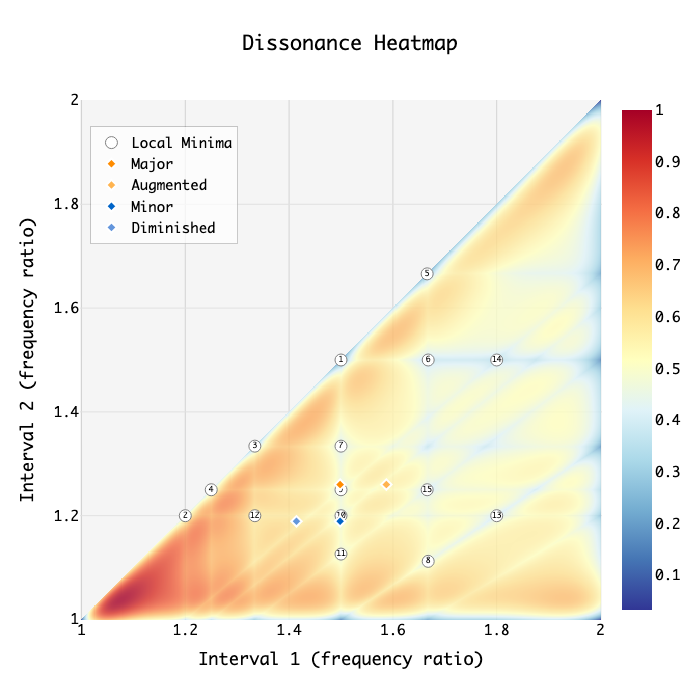}
    \caption{Two-dimensional dissonance surface $D_F(r, s)$ for triads, evaluated over all frequency ratio pairs $(r, s) \in [1, 2] \times [1, 2]$. Blue regions are consonant valleys; red ridges are dissonant peaks. Numbered markers indicate local-minima. Diamond markers show the positions of the major, minor, augmented, and diminished triads of 12-TET.}
    \label{fig:2d_dissonance}
\end{figure}

\begin{figure}[H]
    \centering
    \includegraphics[width=\columnwidth]{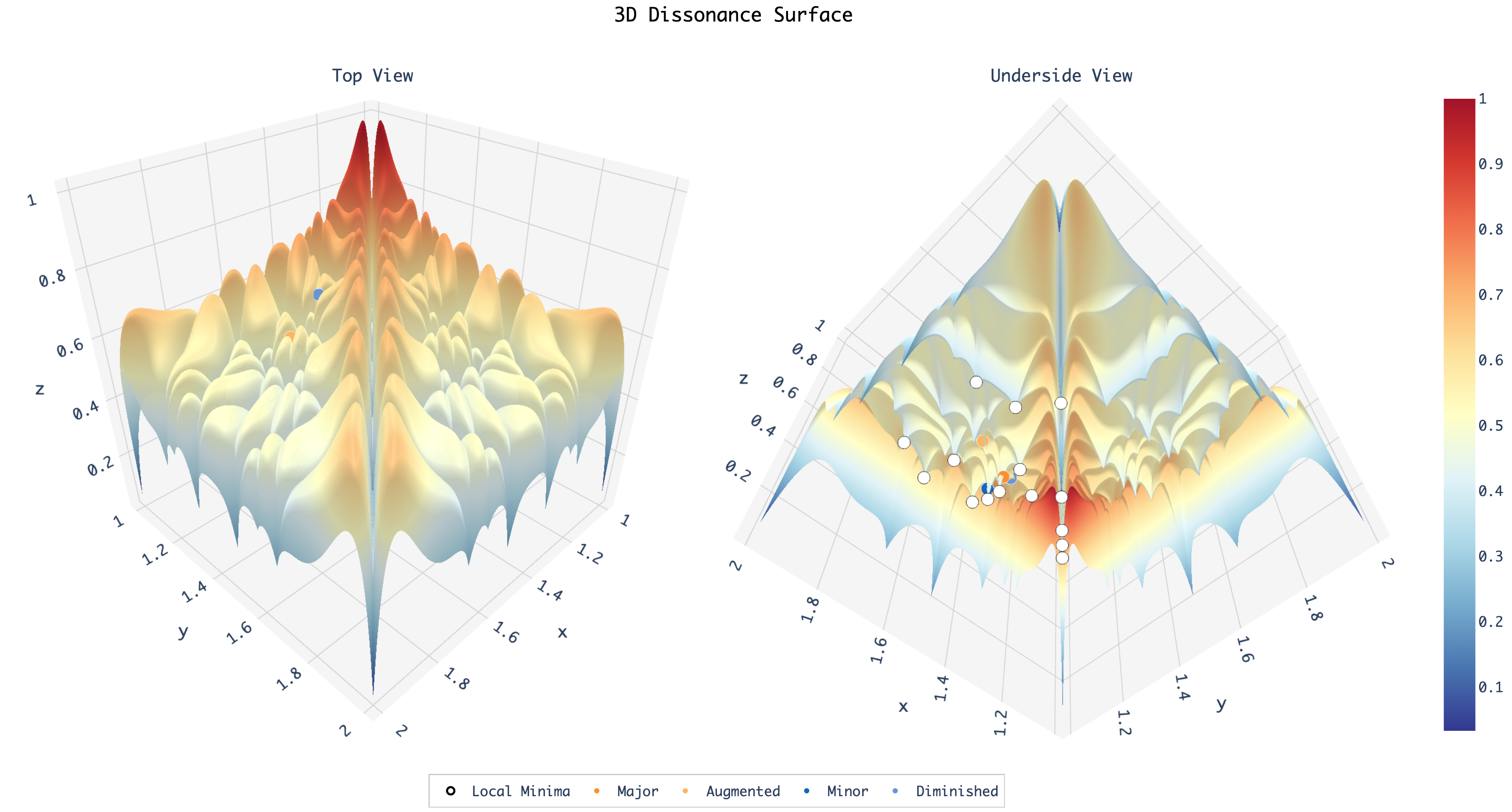}
    \caption{Three-dimensional dissonance surface $D_F(r, s)$ for triads. Consonant valleys contain the local minima (white markers), the most consonant triads of the surface. Coloured markers show the positions of the major, minor, augmented, and diminished triads of 12-TET.}
    \label{fig:3d_dissonance}
\end{figure}

With three simultaneous notes, the root is fixed, and the two upper voices are free to vary. Their positions are described by two frequency ratios $r$ and $s$ relative to the root, both ranging across one octave: $r, s \in [1, 2]$. Evaluating $D_F(r, s)$ at every point on this grid produces a two-dimensional dissonance surface, visualised as a heatmap in Figure~\ref{fig:2d_dissonance} and as a three-dimensional surface in Figure~\ref{fig:3d_dissonance}.

The surface is the evaluation of $D_F(r, s)$ from Equation~\ref{eq:chord_dissonance} across all pairs $(r, s) \in [1, 2] \times [1, 2]$, with the constraint $s \leq r$ to restrict the domain to unique triads.

Because swapping $r$ and $s$ yields the same chord, only the region where $s \leq r$ contains unique triads. The surface is therefore restricted to this lower triangle.

The valleys contain isolated local-minima, where the partials of all three notes interact least. The major, minor, augmented, and diminished triads of 12-TET each map to a specific $(r, s)$ coordinate determined by their interval ratios, and these coordinates fall near, but not exactly at, the local-minima of the surface. Every chord type in any tuning system maps to a unique point on this surface, positioned entirely by its interval structure and the resulting psychoacoustic dissonance.

% \begin{figure}[t]
%     \centering
%     \includegraphics[width=\columnwidth]{Figures/3D_Dissonance_up.png}
%     \caption{}
%     \label{fig:3d_dissonance_up}
% \end{figure}

\subsection{The Harmonic Eigenspace}
\label{subsec:Eigenspace}

Adding a fourth note introduces a third free interval. With the root fixed as the reference point, the three upper voices are parameterised by frequency ratios $\alpha$, $\beta$, and $\gamma$ relative to the root, all within one octave: $\alpha, \beta, \gamma \in [1, 2]$. The frequency vector of a tetrad rooted at $f$ is:
\begin{equation}
\mathbf{v} = f \cdot (1,\, \alpha,\, \beta,\, \gamma)^{\top}
\label{eq:chord_vector}
\end{equation}
For four simultaneous notes at intervals $1$, $\alpha$, $\beta$, $\gamma$ relative to the root, the total dissonance is the sum of pairwise interactions across all six pairs of voices:
\begin{equation}
D(\alpha, \beta, \gamma) = D_F(\alpha) + D_F(\beta) + D_F(\gamma) + D_{\alpha F}\!\left(\frac{\beta}{\alpha}\right) + D_{\alpha F}\!\left(\frac{\gamma}{\alpha}\right) + D_{\beta F}\!\left(\frac{\gamma}{\beta}\right)
\label{eq:tetrad_dissonance}
\end{equation}
\noindent where each term is evaluated using Equation~\ref{eq:dissonance_curve}. The level sets of this function define the four-dimensional dissonance volume.

\begin{figure}[H]
    \centering
    \includegraphics[width=\columnwidth]{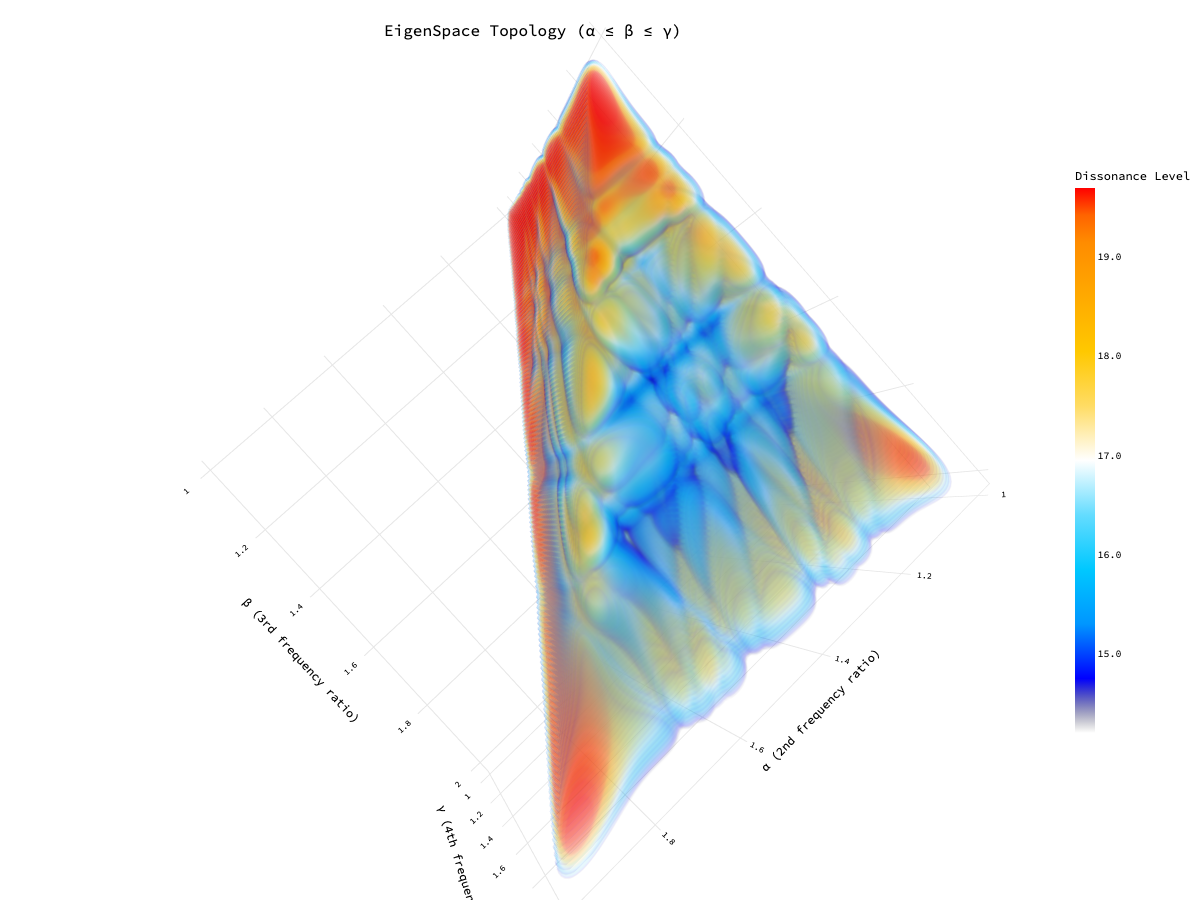}
    \caption{The dissonance volume $D(\alpha, \beta, \gamma)$ rendered over the tetrahedral domain $1 \leq \alpha \leq \beta \leq \gamma \leq 2$ (the ordering removes permutational duplicates), colour-coded from consonant (blue) to dissonant (red). The volume appears as a cloud of regions because the field is continuous: every point is a sounding four-note combination, defined by its three interval ratios above the root and belonging to no particular tuning system, and the application can audition any of them. Dissonant regions concentrate where adjacent voices come close together; the blue pockets are the consonance regions whose deepest points are the local minima. The placement of the 12-, 31-, and 53-TET chord vocabularies inside this continuum is shown in Figures~\ref{fig:chords_mapping} and~\ref{fig:chords_mapping_2}.}
    \label{fig:4d_dissonance}
\end{figure}

The name \emph{Eigenspace} refers to the following structure. Transposition acts on pitch space as the uniform scaling $T_\lambda = \lambda I$: moving a chord from root $f$ to root $\lambda f$ replaces $\mathbf{v}$ with $\lambda\mathbf{v}$ and leaves every interval ratio unchanged. All transpositions of a chord therefore lie on one ray through the origin, and each ray corresponds to one chord quality: for instance, all $\mathrm{maj7}$ ($C\,\mathrm{maj7}$, $D\,\mathrm{maj7}$, $F\sharp\,\mathrm{maj7}$, etc.) lie on the same ray. Since $T_\lambda$ is a scalar operator, every vector is trivially an eigenvector of it; the name therefore does not designate an eigenspace in the strict linear-algebraic sense but the space of invariant directions (eigendirections). Chord quality is a direction, the scale factor $\lambda$ records the transposition interval, and $(\alpha, \beta, \gamma)$, the representative of a ray with root $1$, is a global coordinate chart on the space of qualities.

Note that even though the dissonance value attached to a point does depend on the register, the locations of the local minima do not inherit this dependence: a minimum sits where partials of the sounding tones coincide, and coincidence is a condition on frequency ratios alone, so the coordinates of a minimum are the same at every register
%The dissonance value attached to a point does depend on the register: the scaling factor of Equation~\ref{eq:scaling} depends on the frequencies themselves, not only on their ratios, through the critical-bandwidth term $s_1 f_1 + s_2$, so the same interval ratios produce more roughness in a low register than in a high one, in line with the perceptual measurements the function was fitted to. The locations of the local minima do not inherit this dependence: a minimum sits where partials of the sounding tones coincide, and coincidence is a condition on frequency ratios alone, so the coordinates of a minimum are the same at every register; register determines only whether a shallow well survives against the surrounding roughness. The Eigenspace can therefore be computed at any register.

Each point in the Eigenspace corresponds to the closed-position of a chord type, with all upper voices within one octave of the root.

%Octave displacements of individual voices (voicings) are realisations of a type, not points of the space, and are handled in the application: in the Eigenspace view any voice can be octave-doubled and re-auditioned, and the \textit{Modal Studio} provides a dedicated Voicing Editor for distributing chord tones across registers (Section~\ref{sec:application}). The local minima of the volume are its most consonant tetrads: small perturbations in any interval ratio increase the total dissonance.}

Specific tuning systems appear as discrete lattices within this volume. The 53-TET grid samples the Eigenspace at intervals of $2^{k/53}$, closely approximating many just intonation ratios. The 12-TET grid is a coarser subset, sampling at intervals of $2^{k/12}$. A dominant seventh chord in 12-TET, a septimal dominant seventh in just intonation (with the $7{:}4$ minor seventh), and their nearest 53-TET approximations occupy distinct but neighbouring positions, and their dissonance values account for the perceptual differences between these chords. Qualities from any tuning system coexist in the same space and can be compared on the same psychoacoustic basis; Figures~\ref{fig:chords_mapping} and~\ref{fig:chords_mapping_2} show the lattices inside the volume. 

% --------------------------------------------------------------------------------------------
\section{Harmonic Extension Principle}
\label{sec:harmonic_extension}

The Eigenspace locates chord types by their psychoacoustic properties; this section addresses how a composer moves between them. Modal harmony organises a progression around the characteristic colour of each scale degree rather than around dominant--tonic syntax. In 12-TET, modal interchange is a standard reharmonisation technique: chords are borrowed from parallel or relative modes to introduce new harmonic colour into a progression. The interval vocabulary of 31-, 53-TET, with its five and ten gradations of third and seventh quality, extends this principle into territory that 12-TET cannot reach. Modes built from non-standard Holdrian comma (HC) distributions generate chord types absent from the diatonic vocabulary, and interchange between such modes produces substitution possibilities that grow combinatorially with the number of available interval qualities.

\subsection{Chord Quality Extensions}
\label{chord_quality_extensions}

The chord vocabulary embedded in the Eigenspace is constructed systematically for each temperament. In 12-TET, the standard set of 12 tetrad types covers the familiar qualities: maj7, min7, dom7, \o 7, o7, minmaj7, 7sus2, 7sus4, maj6, min6, power, and sus2, the last two realised as four-note dispositions (power as root, fourth, fifth, and octave-doubled root; sus2 sounding all four of root, second, fourth, and fifth); and from there it is possible to add extensions to the chord. In 31-TET, the five gradations of third quality (sub-minor, minor, neutral, major, super-major) are combined with the same five gradations of seventh quality to form a $5\times5$ grid of 25 chords (Table~\ref{tab:31tet}), with three additional chords (\o 7, o7, M+7) for a total of 28. In 53-TET, the ten third gradations paired with ten seventh gradations produce a $10\times10$ main grid of 100 chords (Table~\ref{tab:53tet-main}), with the fifth fixed at step 31 (the closest approximation to the just perfect fifth). Separate grids cover the half-diminished chords with a flattened fifth at step 26 (Table~\ref{tab:53tet-hdim}, 40 chords), augmented chords with a raised fifth at step 35 (Table~\ref{tab:53tet-aug}, 10 chords), and two suspended chords, for a total of 152 (take into account, \textit{dim}, \textit{augmented}, and \textit{sus} chord can also be extended much further). The grand total across all three temperaments is $12 + 28 + 152 = \mathbf{192}$ chords. Each chord is placed in the Eigenspace at the $(\alpha, \beta, \gamma)$ coordinates computed from its interval ratios, and coloured according to its dissonance value.

\begin{table}[H]
\centering
\footnotesize
\caption{31-TET chord names (28 chords). The $5\times5$ grid varies the third (rows) and seventh (columns) quality; the fifth is fixed at step~18. Three additional chords with different fifths are also considered: (\o 7, o7, M+7)}
\label{tab:31tet}
\begin{tabular}{c|ccccc}
\toprule
 & SM & M & n & m & sm \\
\midrule
SM & SMSM7 & SMM7  & SMn7  & SMm7  & SMsm7  \\
M  & MSM7  & Maj7  & Mn7   & Dom7  & Msm7   \\
n  & nSM7  & nM7   & nn7   & nm7   & nsm7   \\
m  & mSM7  & mM7   & mn7   & min7  & msm7   \\
sm & smSM7 & smM7  & smn7  & smm7  & smsm7  \\
\bottomrule
\end{tabular}

\medskip
\textit{Diminished 31-TET}

\begin{tabular}{lll}
\toprule
\o 7 & o7 & M+7 \\
\bottomrule
\end{tabular}
\end{table}

\begin{table}[H]
\centering
\footnotesize
\caption{53-TET main grid chord names (100 chords). The third quality (rows) and seventh quality (columns) are combined as \texttt{\{third\}\{seventh\}7}; the fifth is fixed at step~31.}
\label{tab:53tet-main}
\begin{tabular}{c|cccccccccc}
\toprule
 & SM & \^{}M & maj & vM & N & n & \^{}m & m & vm & sm \\
\midrule
SM  & SMSM7  & SM\^{}M7  & SMmaj7  & SMvM7  & SMN7  & SMn7  & SM\^{}m7  & SMm7  & SMvm7  & SMsm7  \\
\^{}M  & \^{}MSM7  & \^{}M\^{}M7  & \^{}Mmaj7  & \^{}MvM7  & \^{}MN7  & \^{}Mn7  & \^{}M\^{}m7  & \^{}Mm7  & \^{}Mvm7  & \^{}Msm7  \\
M   & MSM7   & M\^{}M7   & Mmaj7   & MvM7   & MN7   & Mn7   & M\^{}m7   & Mm7   & Mvm7   & Msm7   \\
vM  & vMSM7  & vM\^{}M7  & vMmaj7  & vMvM7  & vMN7  & vMn7  & vM\^{}m7  & vMm7  & vMvm7  & vMsm7  \\
N   & NSM7   & N\^{}M7   & Nmaj7   & NvM7   & NN7   & Nn7   & N\^{}m7   & Nm7   & Nvm7   & Nsm7   \\
n   & nSM7   & n\^{}M7   & nmaj7   & nvM7   & nN7   & nn7   & n\^{}m7   & nm7   & nvm7   & nsm7   \\
\^{}m  & \^{}mSM7  & \^{}m\^{}M7  & \^{}mmaj7  & \^{}mvM7  & \^{}mN7  & \^{}mn7  & \^{}m\^{}m7  & \^{}mm7  & \^{}mvm7  & \^{}msm7  \\
m   & mSM7   & m\^{}M7   & mmaj7   & mvM7   & mN7   & mn7   & m\^{}m7   & mm7   & mvm7   & msm7   \\
vm  & vmSM7  & vm\^{}M7  & vmmaj7  & vmvM7  & vmN7  & vmn7  & vm\^{}m7  & vmm7  & vmvm7  & vmsm7  \\
sm  & smSM7  & sm\^{}M7  & smmaj7  & smvM7  & smN7  & smn7  & sm\^{}m7  & smm7  & smvm7  & smsm7  \\
\bottomrule
\end{tabular}
\end{table}

\begin{table}[H]
\footnotesize
\centering
\caption{53-TET half-diminished chord names (40 chords). The third quality (rows) and seventh quality (columns) are combined as \texttt{\o\{third\}\{seventh\}7}; the fifth is fixed at step~26.}
\label{tab:53tet-hdim}

\medskip
\textit{Major sevenths (20 chords)}\\
\begin{tabular}{l|lllll}
\toprule
 & SM & \^{}M & maj & vM & N \\
\midrule
\^{}m & \o\^{}mSM7 & \o\^{}m\^{}M7 & \o\^{}mmaj7 & \o\^{}mvM7 & \o\^{}mN7 \\
m  & \o mSM7  & \o m\^{}M7  & \o mmaj7  & \o mvM7  & \o mN7  \\
vm & \o vmSM7 & \o vm\^{}M7 & \o vmmaj7 & \o vmvM7 & \o vmN7 \\
sm & \o smSM7 & \o sm\^{}M7 & \o smmaj7 & \o smvM7 & \o smN7 \\
\bottomrule
\end{tabular}

\medskip
\textit{Minor sevenths (20 chords)}\\

\begin{tabular}{l|lllll}
\toprule
 & n & \^{}m & m & vm & sm \\
\midrule
\^{}m & \o\^{}mn7 & \o\^{}m\^{}m7 & \o\^{}mm7 & \o\^{}mvm7 & \o\^{}msm7 \\
m  & \o mn7  & \o m\^{}m7  & \o mm7  & \o mvm7  & \o msm7  \\
vm & \o vmn7 & \o vm\^{}m7 & \o vmm7 & \o vmvm7 & \o vmsm7 \\
sm & \o smn7 & \o sm\^{}m7 & \o smm7 & \o smvm7 & \o smsm7 \\
\bottomrule
\end{tabular}
\end{table}

\begin{table}[H]
\centering
\footnotesize
\caption{53-TET augmented chord names (10 chords). The third is fixed at step~18 and the fifth at step~35; only the seventh varies.}
\label{tab:53tet-aug}

\medskip
\textit{Major sevenths (5 chords)}

\begin{tabular}{ccccc}
M+SM7 & M+\^{}M7 & M+maj7 & M+vM7 & M+N7 \\
\bottomrule
\end{tabular}

\medskip
\textit{Minor sevenths (5 chords)}

\begin{tabular}{ccccc}
M+n7 & M+\^{}m7 & M+m7 & M+vm7 & M+sm7 \\
\bottomrule
\end{tabular}
\end{table}

\subsection{Modal Interchange in 53-TET}
\label{modal-interchange}
\begin{figure}[H]
  \centering
  \includegraphics[width=0.7\linewidth]{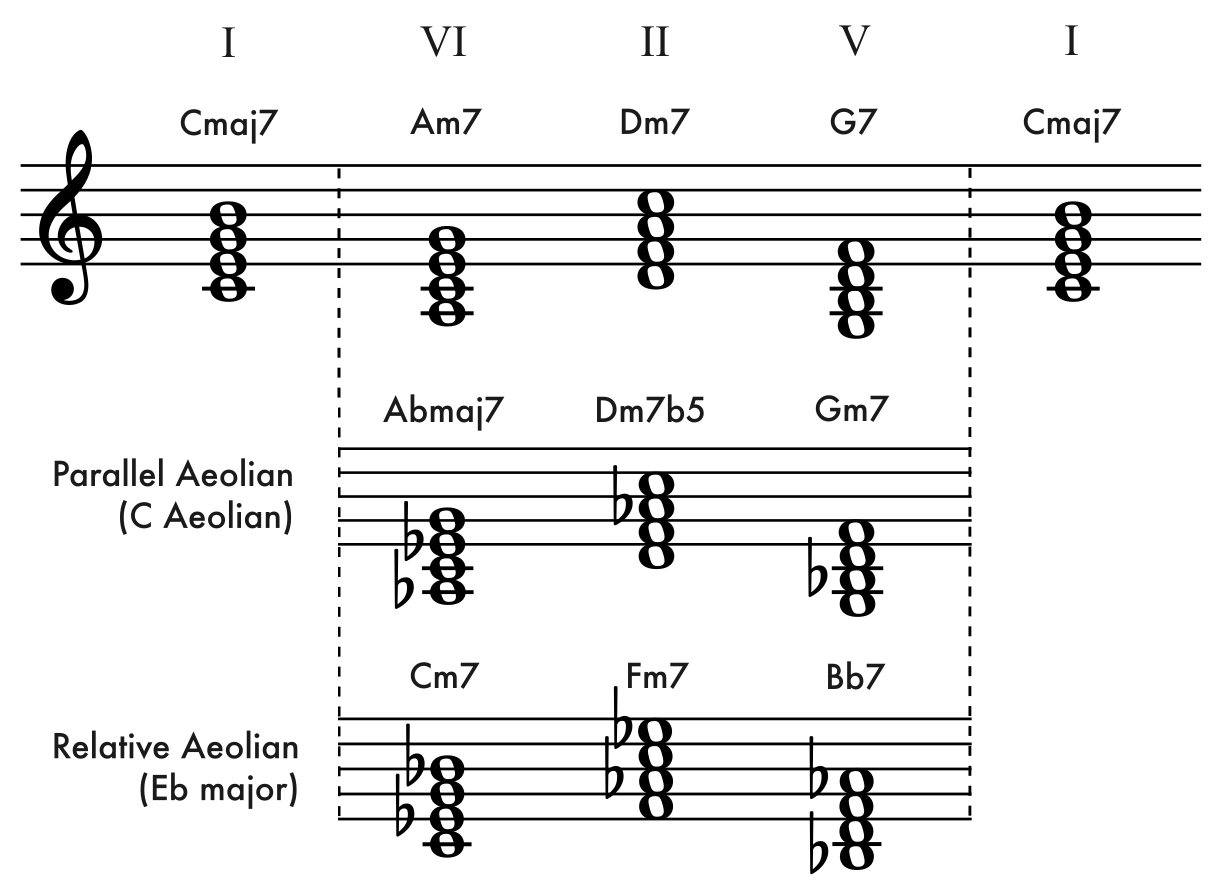}
  \caption{Parallel and relative Aeolian interchanges applied to a \texttt{I vi ii V I} progression in C major. The parallel interchange replaces chords with their C Aeolian counterparts; the relative interchange borrows from E$\flat$ major, where C functions as the 6th degree.}
\label{fig:interchanges}
\end{figure}

Modal interchange introduces chords from outside the current mode into a progression. Two types of interchange are used (see Figure~\ref{fig:interchanges}).

In a \textbf{parallel interchange}, chords are borrowed from a mode that shares the same root. Starting in C Ionian, a parallel Aeolian interchange replaces diatonic chords with their Aeolian counterparts: the flattened 3rd, 6th, and 7th degrees alter the available chord qualities while the tonal centre remains fixed. A progression \texttt{I vi ii V I} in C major yields, under parallel Aeolian interchange, A$\flat$maj7, Dm7$\flat$5, and Gm7 for the middle three chords.

In a \textbf{relative interchange}, the current root is reinterpreted as a different scale degree of another key, and chords are borrowed from that key. Starting again in C Ionian, an Aeolian relative interchange places C as the 6th degree of E$\flat$ major. The middle chords of the same progression become Cm7, Fm7, and B$\flat$7.

In 12-TET, each mode produces a fixed set of seven chord qualities per interchange. In 53-TET, the ten gradations of third quality and ten of seventh quality multiply the number of available substitutions: a single parallel or relative interchange can yield chord types that have no 12-TET equivalent. In Section \ref{sec:application}, we explain how the \textit{Modal Studio} implements both interchange types across all seven diatonic modes, populating the \textit{Grid} with the full set of substitution options for any user-defined scale.

The modal interchange mechanism takes the seven diatonic modes as its starting point. In 53-TET, the standard Ionian mode corresponds to the HC distribution $[9, 9, 4, 9, 9, 9, 4]$, which lands on the same pitch classes as 12-TET (see Figure~\ref{fig:scaleEditor}). Any modification to these step sizes produces a mode with no 12-TET equivalent: for instance, the distribution $[8, 7, 7, 9, 5, 10, 7]$ introduces neutral intervals that generate chord qualities absent from the standard diatonic vocabulary. Each such mode defines its own set of consonance and dissonance relations, and modal interchange between these non-standard modes opens a combinatorial space of substitutions unavailable in 12-TET.

% ------------------------------------------------------------------------------
\section{The Application}
\label{sec:application}
Based on the two principles explained, the computer application provides two areas of exploration: a) The \textit{Harmonic Eigenspace} and b) the \textit{Modal Studio}. The two areas divide the work. The Eigenspace is a sonification of the dissonance volume for positional comparison of the 12-, 31-, and 53-TET vocabularies: a place to survey and audition chord qualities. Composition proceeds in the \textit{Modal Studio}, where scales, progressions, and voicing are developed explicitly. 

\subsection{The Harmonic Eigenspace}
\begin{figure}[H]
    \centering
    \includegraphics[width=1.0\columnwidth]{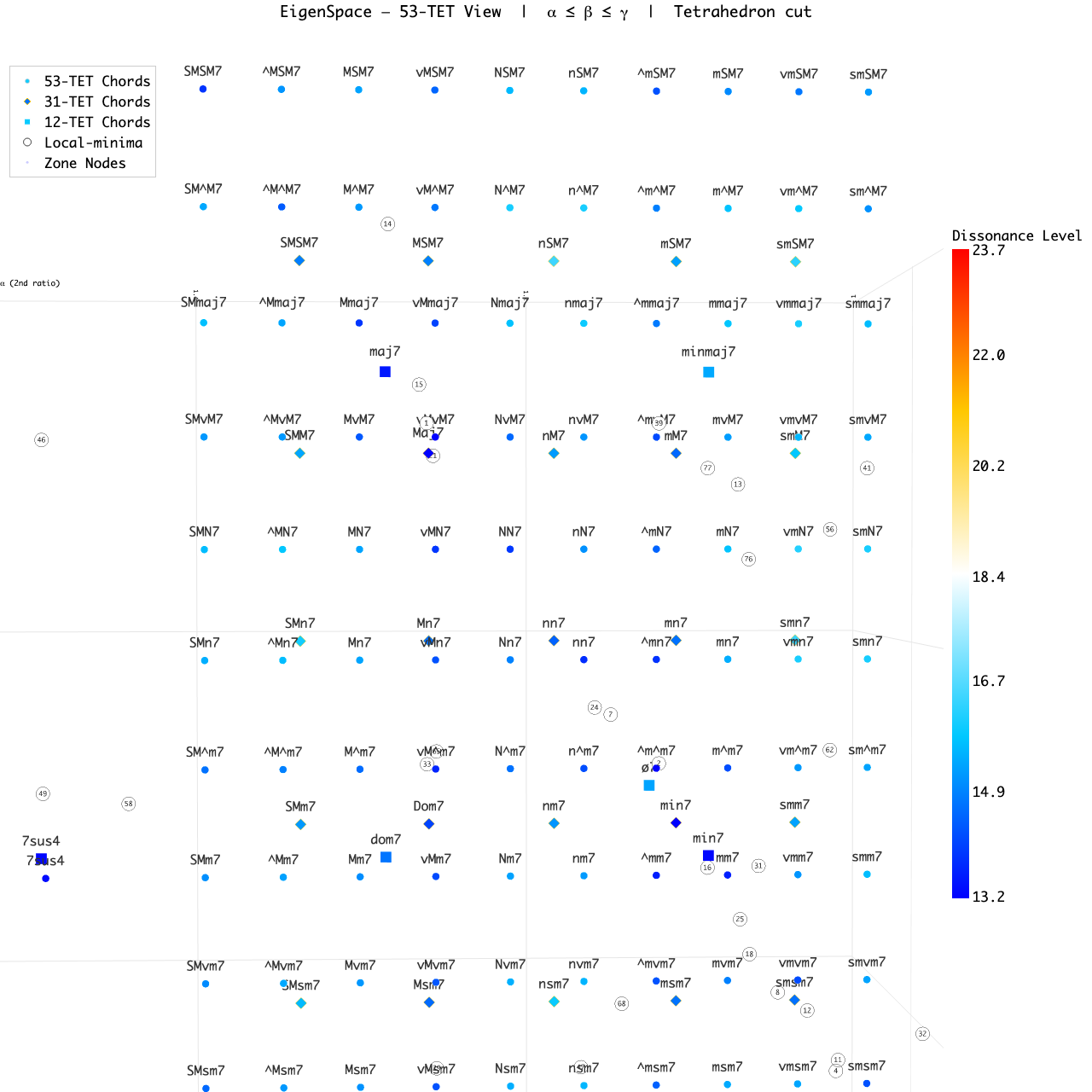}
    \caption{The 12-, 31-, and 53-TET chord positions with zone-nodes hidden, showing the temperament lattices and their dissonance colour-coding in isolation.}
    \label{fig:chords_mapping_2}
\end{figure}

We used the Plotly JavaScript open-source graphing library\footnote{\url{https://plotly.com/javascript/}} to form an interactive visualisation of the Eigenspace. The application is available at \url{https://dazzid.github.io/ANIMA_Harmonic_Eigenspace/}. The visualisation is composed of three types of nodes: 
\begin{enumerate}[label=\alph*)]
    \item The zone-nodes, derived from the 4D dissonance calculation, where each node is clickable and reproduces the sound of the four-note groups. Those groups can be chords or clusters; the space is not limited to any particular temperament. Each node is coloured according to the dissonance mapping, and the visualisation can be navigated as a full cluster of nodes or filtered by degree of dissonance using the `\textit{Switch to Layered View}' button. All dissonance values reported in this paper (the volume of Figure~\ref{fig:4d_dissonance} and the values of Table~\ref{tab:lm_distances}) are computed for tones of six equal-loudness partials, with the root taken at A3 (220~Hz), the sounding root of the application; the root is a starting point for the computation, and any other choice changes only the depths of the field, not its coordinates or the locations of its minima. 

    \item Local-minima (LM) nodes mark the less dissonant points in a region, as shown in Figures~\ref{fig:3d_dissonance}, \ref{fig:chords_mapping_2}. These nodes serve as spatial landmarks, providing clustering distances and locality for each node. For instance, local-minimum 1 corresponds to a just-intonation major seventh, and local-minimum 3 is located close to the \^{}m\^{}m7. A tempered chord appearing nearby in the space can be understood as a tempered variant of that same harmonic region. Because LMs are derived from the continuous dissonance landscape, they are not tempered and align with just-intonation intervals. LMs were identified by evaluating $D(\alpha,\beta,\gamma)$ on a $500^{3}$ grid (each axis spanning the interval $[1,2]$ in steps of $\approx 0.002$) and selecting points whose dissonance value was lower than all neighbours within an $11^{3}$ voxel footprint. Candidates near the domain boundary (within a 50-step margin), with near-degenerate intervals ($\alpha \approx \beta$ or $\beta \approx \gamma$ within two grid steps), or with prominence below 0.001 were discarded. The remaining candidates were ranked by a curvature score (prominence weighted by local gradient magnitude), and the top 100 were retained. Each was then refined by 100 iterations of stochastic perturbation to converge on the true continuous minimum. The number of nodes used is a parameter set to 78 in the web-app from those 100 LM.

    \item Chord positions for 192 chords across three temperaments (12-TET, 31-TET, and 53-TET) are displayed, with each node coloured according to its dissonance value (see Section~\ref{chord_quality_extensions} for the full chord vocabulary). As an example, consider the 12-TET major seventh chord (maj7). Its three frequency ratios relative to the root are $\alpha = 2^{4/12} \approx 1.2599$ (major third), $\beta = 2^{7/12} \approx 1.4983$ (perfect fifth), and $\gamma = 2^{11/12} \approx 1.8877$ (major seventh). For a root at 220\,Hz (A3), the tetrad is given in Table \ref{tab:maj7-example}.
    The closest local minimum (node~1) sits at the just-intonation ratios $\bigl(\tfrac{5}{4},\;\tfrac{3}{2},\;\tfrac{15}{8}\bigr) = (1.2500,\;1.5000,\;1.8750)$. The Euclidean distance from the 12-TET point to that node is:\\

    \(d = \sqrt{(1.2599-1.25)^{2} + (1.4983-1.5)^{2} + (1.8877-1.875)^{2}} \approx 0.016\)\\
    
    confirming that the 12-TET major seventh lies very close to the just-intonation well, near a stable consonance valley.
\end{enumerate}

\begin{table}[t]
\centering
\footnotesize
\caption{12-TET maj7 chord example at 220\,Hz.}
\label{tab:maj7-example}
\begin{tabular}{l c c c}
\toprule
\textbf{Note} & \textbf{Axis} & \textbf{Ratio} & \textbf{Frequency} \\
\midrule
Root (A3)               & ---                          & 1.0000 & 220.0\,Hz \\
Major 3rd (C$\sharp$4)  & $\alpha = 2^{4/12}$          & 1.2599 & 277.2\,Hz \\
Perfect 5th (E4)        & $\beta  = 2^{7/12}$          & 1.4983 & 329.6\,Hz \\
Major 7th (G$\sharp$4)  & $\gamma = 2^{11/12}$         & 1.8877 & 415.3\,Hz \\
\bottomrule
\end{tabular}
\end{table}

\begin{figure}[H]
    \centering
    \includegraphics[width=1.0\columnwidth]{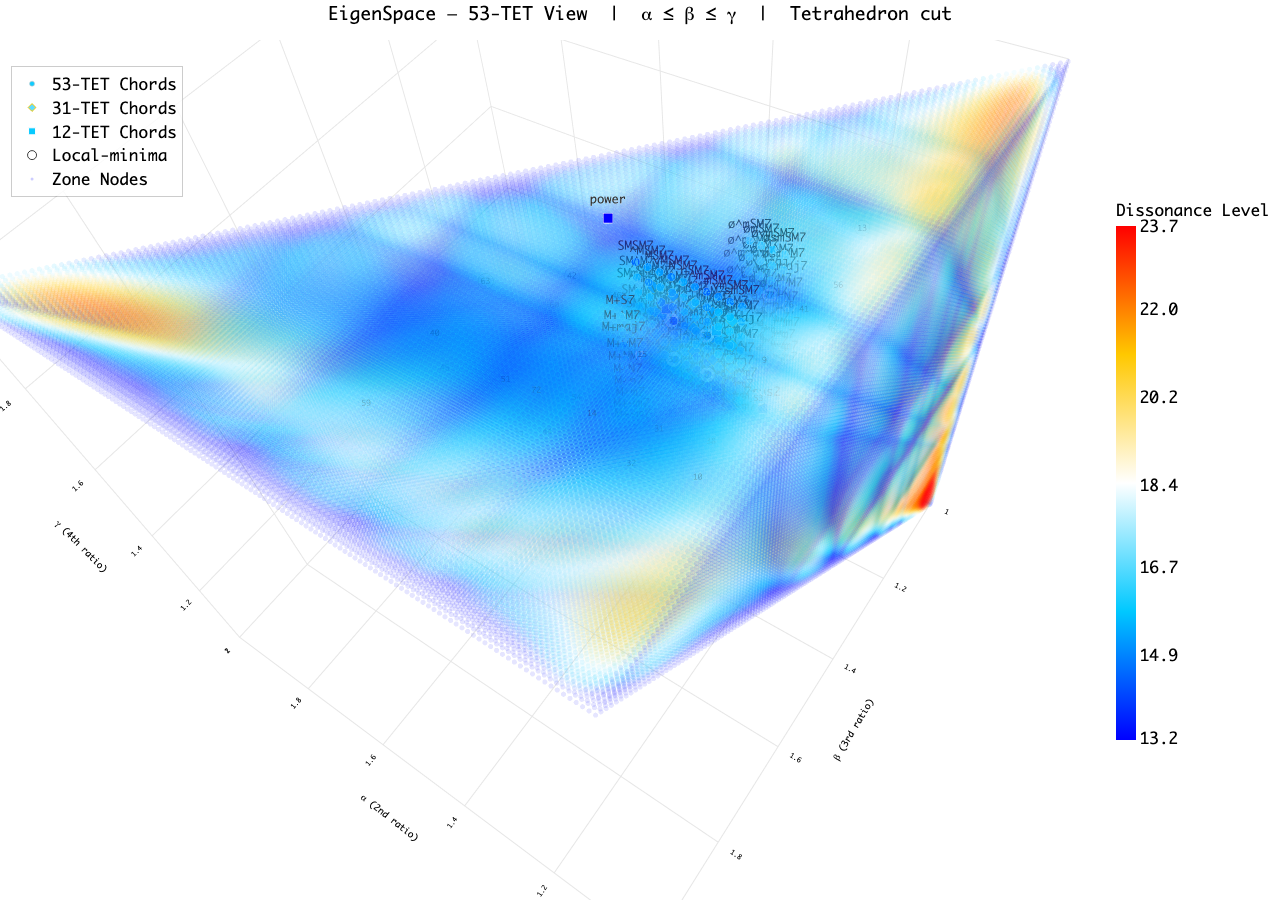}
    \caption{Overview of the zone-nodes with dissonance colour-coding. The 12-, 31-, and 53-TET chords are embedded within the space. A consonance canal is visible along the region where the fifth is perfect; diminished and augmented chords fall into adjacent higher-dissonance regions.}
    \label{fig:chords_mapping}
\end{figure}

Figure~\ref{fig:chords_mapping} shows part of the 12-, 31-, 53-TET chords located inside the Eigenspace, where the colour-code shows the dissonance scale range. Local-minima nodes are drawn in light grey, and zone-nodes are shown with reduced opacity to allow clear visualisation of the tetrad spatial distribution. 

Operational details of the interface (the piano-roll spectrum visualisation, the audio-engine settings, and the app-wide Chord Memory grid) are documented in Appendix~\ref{app:usage}.

\subsection{Modal Studio}

\begin{figure}[H]
    \centering
    \includegraphics[width=0.9\columnwidth]{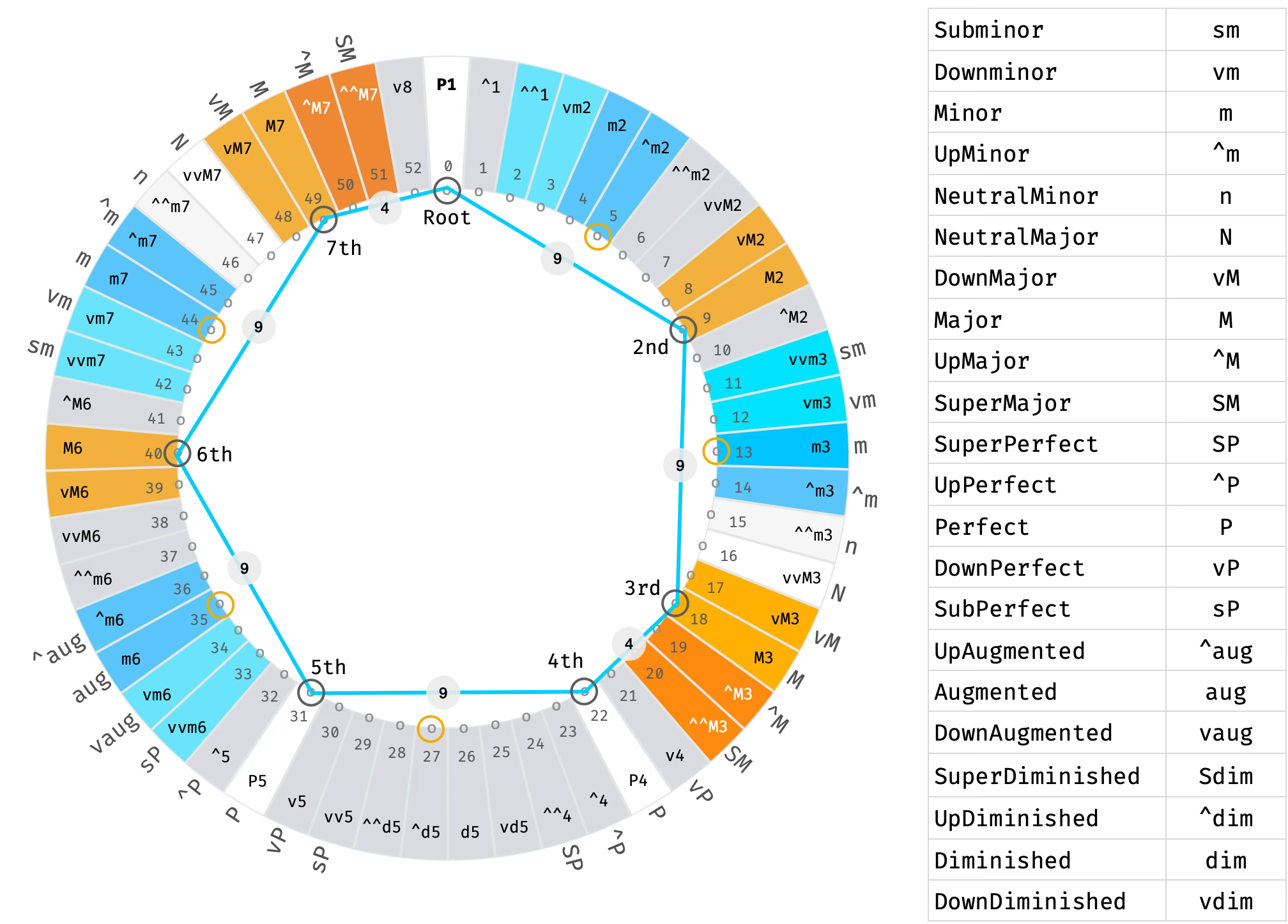}
    \caption{The Scale Editor. A circular visualisation of the 53-TET octave with twelve draggable nodes representing scale degrees. Repositioning a node redefines the interval pattern and triggers automatic derivation of all seven diatonic modes and their chord qualities.}
    \label{fig:scaleEditor}
\end{figure}

The \textit{Modal Studio} is an interactive compositional environment for exploring microtonal harmony in 53-TET through the lens of modal interchange. At its core, the system comprises three interconnected editors. The Scale Editor (Figure~\ref{fig:scaleEditor}) presents a circular visualisation of the 53-TET octave with twelve draggable nodes on its outer ring, each representing a scale degree. By repositioning these nodes, the user reshapes the underlying interval pattern, and the system automatically derives all seven diatonic modes (Ionian through Locrian) along with their associated chord qualities. 

\begin{figure}[H]
    \centering
    \includegraphics[width=0.9\columnwidth]{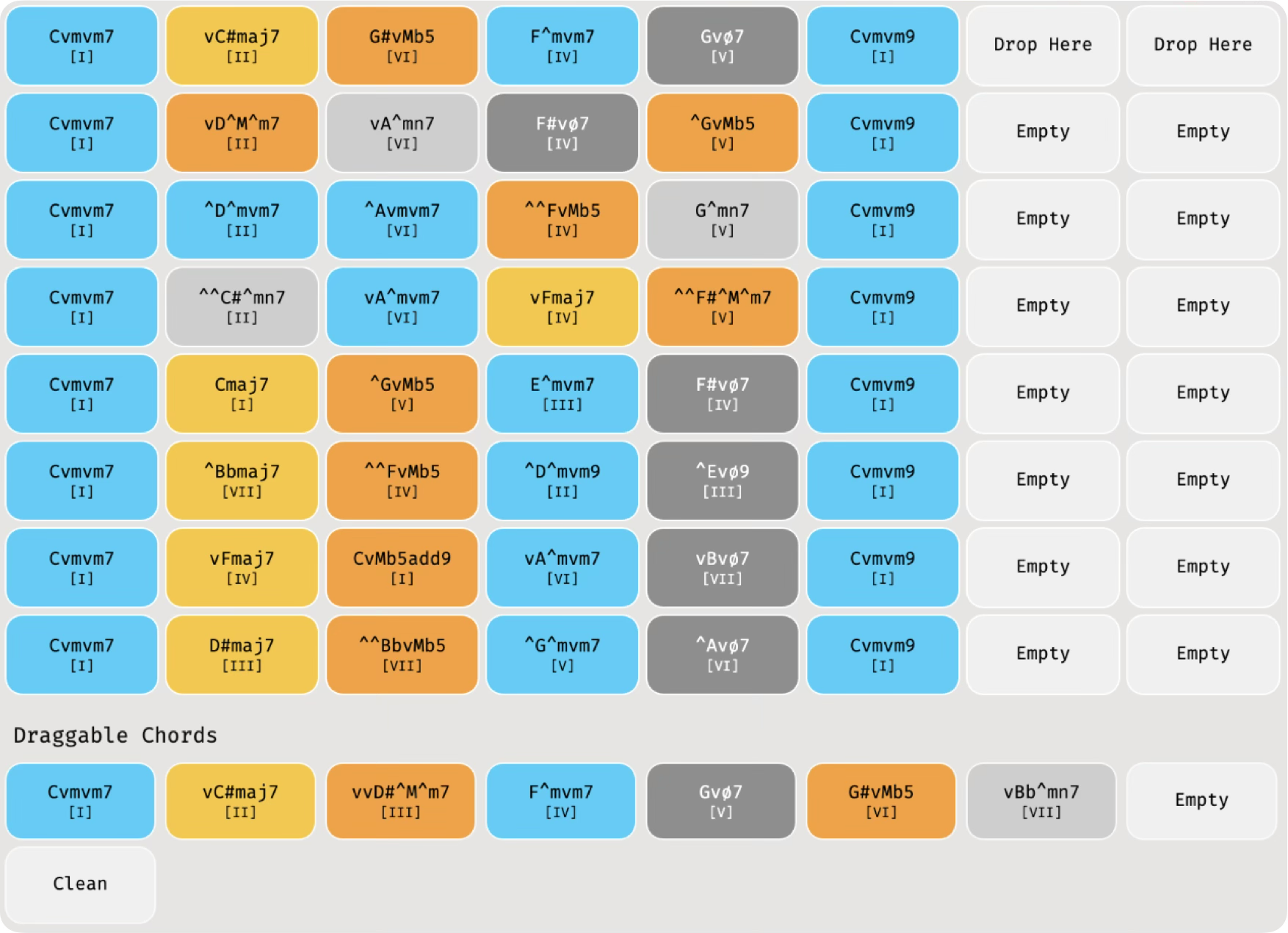}
    \caption{The Grid. An 8$\times$8 chord-progression sequencer. The first row holds the user's primary progression; the remaining rows are populated automatically with parallel and relative modal substitutions.}
    \label{fig:modal_grid}
\end{figure}

Once a scale is defined, chords built on each degree can be dragged from a side panel into the first row of the \textbf{Grid} (Figure~\ref{fig:modal_grid}), an 8$\times$8 matrix that serves as a chord-progression sequencer. The first row holds the user's primary progression; the remaining seven rows are populated automatically with modal substitutions, including parallel modes (same root, different modal quality) and relative modes (different root, same intervallic structure), enabling the composer to audition modal interchange options at a glance. This is explained further in section \ref{modal-interchange}.

\begin{figure}[H]
    \centering
    \includegraphics[width=0.6\columnwidth]{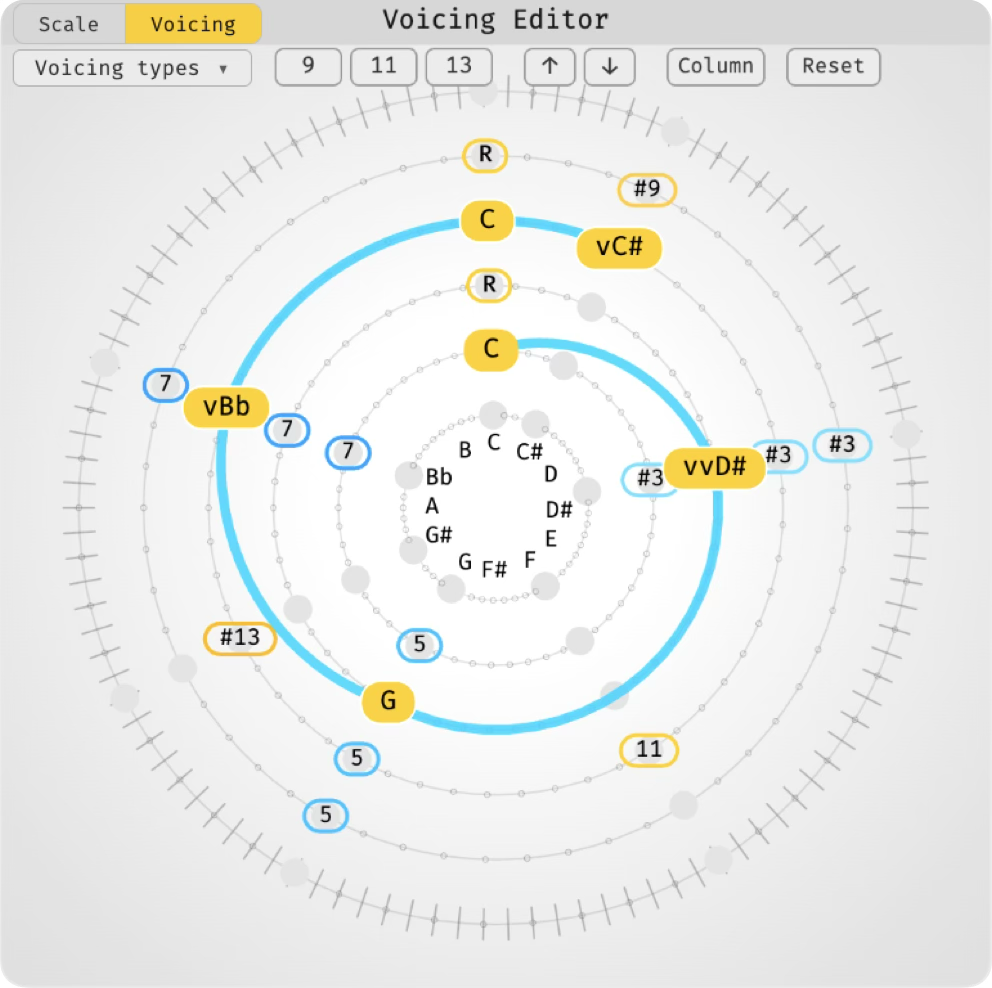}
    \caption{The Voicing Editor. A concentric spiral display spanning five octaves in which individual chord tones can be repositioned to craft specific distributions across registers.}
    \label{fig:voicingEditor}
\end{figure}

Selecting any cell in the Grid triggers audio playback via the audio engine and opens the chord in the Voicing Editor (Figure~\ref{fig:voicingEditor}). Following \citet{chew2007out}, individual notes can be dragged to craft specific voicings by placing them on a concentric spiral display spanning five octaves. Voice-leading continuity is preserved across substitutions: when a chord is replaced by a modal variant, the system selects from a list of voicing formats the closest consecutive disposition to ensure a smooth progression. The 9th extension is used as a pivotal note that can be added to maintain voice leading. 

% --------------------------------------------------------------------------------------------
\section{Listening Test}
\label{sec:listening_test}

A listening study was conducted to address two questions: how listeners rate the dissonance of 53-TET chords drawn from the application's Eigenspace, and how tolerant listeners are to microtonal chord progressions composed using the \textit{Modal Interchange} mode of the web-app. In total, 31 participants (22 male, 5 female, 2 prefer not to say, 2 non-binary) completed a listening study using jsPsych \citep{de2023jspsych}. Their mean age was 36.0 years (SD = 10.4, range 22–58). The average completion time was 17.7 minutes (SD = 23.1). The study consisted of two phases. Following informed consent (age, gender, microtonal experience), participants completed a preliminary harmonic identification task requiring them to identify chord progressions from audio examples. In Part 1, participants rated the dissonance of 19 isolated microtonal chords (representing six categories: Consonant, Intermediate, Neutral, Less Consonant, Dominant, and Diminished) on a scale from 1 (very consonant) to 10 (very dissonant), with white noise presented between trials. In Part 2, participants rated 10 chord progressions on four dimensions-harmony coherence, musicality, dissonance/consonance, and novelty, using identical 1–10 scales. Each chord progression was presented with its underlying harmonic notation. Participants were required to listen to complete audio stimuli before responding. All trials were presented in randomised order, with white noise palate cleansers separating progression trials. The experiment included an exit survey soliciting comments and reports of difficulties. Dependent variables included dissonance ratings for individual chords and ratings of harmony, musicality, dissonance, and novelty for progressions.

The dataset, audio stimuli, and experiment code are publicly available.\footnote{\url{https://anonymous.4open.science/r/Harmonic_Eigenspace_Dataset-2086/}}

\subsection{Part 1: Chord Dissonance Ratings}
\label{sec:dissonance_ratings}

In Part 1, the participants rated 20 isolated 53-TET seventh-chord stimuli; a duplicated \textsf{$vM^m7$} presentation was excluded from analysis, leaving 19 distinct chords on a dissonance scale from 1 (very consonant) to 10 (very dissonant). The Eigenspace dissonance values range from 11.587 to 17.806. During the introduction to the task, listeners were familiarised with the ends of the rating scale by two anchor chords: \textsf{maj6}, presented as an example of 1 (very consonant), and \textsf{M+sm7}, presented as an example of 10 (very dissonant). For the comparison between the model and the ratings, the Eigenspace dissonance values of the 19 tested chords were linearly rescaled to the interval range from 1 to 10, inside the anchor-defined ends of the scale; since this rescaling is affine, it has no effect on the correlation analysis, and the anchor chords themselves are not part of the analysed set.

\begin{figure}[H]
  \centering
  \includegraphics[width=1.0\linewidth]{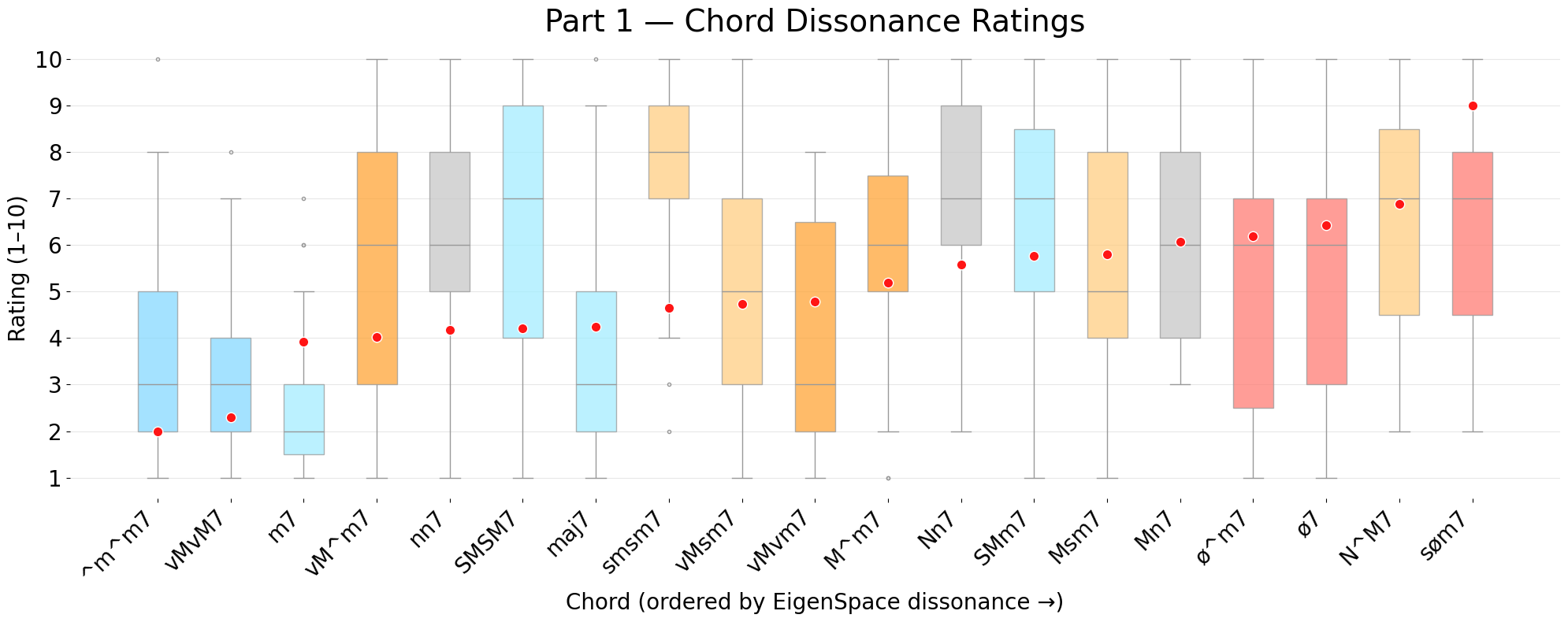}
  \caption{Part 1: Perceived dissonance ratings for 19 isolated 53-TET chords, ordered by Eigenspace dissonance. The boxes show the interquartile range with the median line. Red dots indicate the Eigenspace dissonance value, scaled to the listener-rating range. Box colours indicate chord category: Consonant (light blue), Intermediate (pale blue), Neutral (grey), Dominant (orange), Less Consonant (light-orange), Diminished (red)}
\label{fig:part_1}
\end{figure}

The chord receiving the lowest mean dissonance rating was \textsf{m7} (Intermediate category; M = 2.52, SD = 1.57), while the highest was \textsf{smsm7} (Less Consonant; M = 7.52, SD = 1.98). At the category level, Consonant chords received the lowest mean perceived dissonance (M = 3.37), followed by Intermediate (M = 4.89), Dominant (M = 5.06), Diminished (M = 5.61), Less Consonant (M = 6.16), and Neutral (M = 6.40), as shown in Figure~\ref{fig:part_1}.

Inter-rater agreement, measured by standard deviation, was highest for \textsf{m7} (Intermediate; SD = 1.57) and \textsf{vMvM7} (Consonant; SD = 1.74), suggesting strong consensus on these chords. Agreement was lowest for \textsf{vM\^{}m7} in the Dominant category (SD = 2.79) and \textsf{\o 7} (Diminished; SD = 2.69), indicating greater perceptual ambiguity for these chords.

Table~\ref{tab:lm_distances} lists the Euclidean distance from each of the 19 listening test chords to its closest local minimum in the $(\alpha, \beta, \gamma)$ space. The Consonant chords sit within 0.002 of their nearest minima, Intermediate and Neutral chords at 0.025 to 0.035, Less Consonant chords at 0.05 to 0.07, and Diminished chords at 0.03 to 0.04. 

\begin{table}[ht]
\centering
\footnotesize
\caption{Euclidean distance from each listening test chord to its closest local minimum in ($\alpha$,$\beta$,$\gamma$) space. Chords are grouped by perceptual category.}
\label{tab:lm_distances}
\begin{tabular}{llccccccc}
\toprule
\# & Chord & Category & $\alpha$ & $\beta$ & $\gamma$ & $D_{\text{chord}}$ & Closest LM & Distance \\
\midrule
1 & vMvM7  & Consonant       & 1.2490 & 1.4999 & 1.8734 & 11.855 & LM-1  & 0.0016 \\
2 & \^{}m\^{}m7 & Consonant  & 1.2009 & 1.4999 & 1.8013 & 11.587 & LM-3  & 0.0015 \\
\midrule
3 & SMSM7  & Intermediate    & 1.2990 & 1.4999 & 1.9484 & 13.556 & LM-61 & 0.0805 \\
4 & SMm7   & Intermediate    & 1.2990 & 1.4999 & 1.7779 & 14.932 & LM-63 & 0.0352 \\
5 & maj7   & Intermediate    & 1.2654 & 1.4999 & 1.8981 & 13.584 & LM-1  & 0.0280 \\
6 & m7     & Intermediate    & 1.1853 & 1.4999 & 1.7779 & 13.287 & LM-3  & 0.0271 \\
\midrule
7 & Mn7    & Neutral         & 1.2654 & 1.4999 & 1.8250 & 15.209 & LM-34 & 0.0280 \\
8 & Nn7    & Neutral         & 1.2328 & 1.4999 & 1.8250 & 14.764 & LM-34 & 0.0293 \\
9 & nn7    & Neutral         & 1.2167 & 1.4999 & 1.8250 & 13.526 & LM-3  & 0.0296 \\
\midrule
10 & N\^{}M7  & Less consonant & 1.2328 & 1.4999 & 1.9230 & 15.923 & LM-1  & 0.0514 \\
11 & vMsm7  & Less consonant & 1.2490 & 1.4999 & 1.7320 & 14.022 & LM-7  & 0.0665 \\
12 & Msm7   & Less consonant & 1.2654 & 1.4999 & 1.7320 & 14.959 & LM-7  & 0.0683 \\
13 & smsm7  & Less consonant & 1.1547 & 1.4999 & 1.7320 & 13.943 & LM-10 & 0.0530 \\
\midrule
14 & M\^{}m7  & Dominant      & 1.2654 & 1.4999 & 1.8013 & 14.417 & LM-34 & 0.0152 \\
15 & vM\^{}m7 & Dominant      & 1.2490 & 1.4999 & 1.8013 & 13.380 & LM-34 & 0.0013 \\
16 & vMvm7  & Dominant        & 1.2490 & 1.4999 & 1.7548 & 14.060 & LM-34 & 0.0468 \\
\midrule
17 & o\^{}m7  & Diminished    & 1.2009 & 1.4050 & 1.8013 & 15.309 & LM-8  & 0.0380 \\
18 & som7   & Diminished      & 1.1547 & 1.4050 & 1.7779 & 17.806 & LM-18 & 0.0367 \\
19 & o7     & Diminished      & 1.1853 & 1.4050 & 1.7779 & 15.527 & LM-18 & 0.0426 \\
\bottomrule
\end{tabular}
\end{table}

\subsection{Part 2: Chord Progressions Ratings}
\label{sec:chord_progressions}

\begin{figure}[H]
  \centering
  \includegraphics[width=1.0\linewidth]{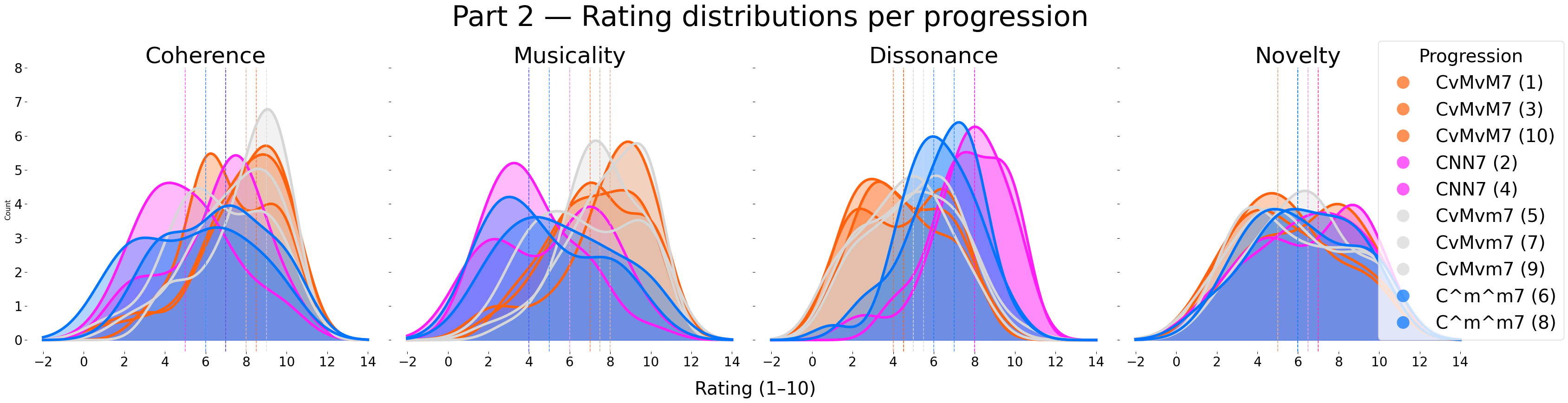}
  \caption{Part 2: Rating distributions per progression across four perceptual dimensions. Progressions are grouped by tonic type: Major (orange), Neutral (purple), Dominant (grey), Minor (blue). Dashed lines indicate medians.}
\label{fig:part_2}
\end{figure}

In Part~2, participants listened to 10 chord progressions composed in 53-TET using the web-app's modal interchange interface. Listeners rated each along four perceptual dimensions on a scale from 1 to~10: \emph{coherence}, \emph{musicality}, \emph{dissonance}, and \emph{novelty}.
Each progression begins and ends on the same tonic chord, contains between five and seven seventh chords, and was generated using voice-leading rules within the Eigenspace harmonic framework.
Progressions were grouped into four chord types related to the concepts of the quality of their tonic chord: \textbf{Major} (\textsf{vMvM7}, progressions 1, 3, 10, in orange), \textbf{Neutral} (\textsf{NN7}, progressions 2, 4, in purple), \textbf{Minor} (\textsf{\^{}m\^{}m7}, progressions 6, 8, in blue), and \textbf{Dominant} (\textsf{vMvm7}, progressions 5, 7, 9, in grey), shown in Figure~\ref{fig:part_2}.

Table~\ref{tab:progressions} lists the chord progressions:
\begin{table}[H]
\centering
\scriptsize
\caption{Chord progressions used in Part~2 of the listening test, grouped by tonic type.}
\label{tab:progressions}
\begin{tabular}{c l l l l l l}
\toprule
\textbf{\#} & \textbf{Step 1} & \textbf{Step 2} & \textbf{Step 3} & \textbf{Step 4} & \textbf{Step 5} & \textbf{Step 6} \\
\midrule
\multicolumn{7}{l}{\textit{Major tonic} (\textsf{vMvM7})} \\
\midrule
1  & CvMvM7[I]       & G\#vMvM7/C[VI]     & C\#vMvM7[II]       & F\^{}m\^{}m7[IV]       & D\#vMm7[III]        & CvMvM7[I]       \\
3  & CvMvM7[I]       & C\#vMvM7/F[II]     & F\#vMvM7[V]        & Bb\^{}m\^{}m7[VII]     & G\#vMm7[VI]         & CvMvM7[I]       \\
10 & CvMvM7[I]       & D\#vMvM7/G[III]    & G\#vMvM7[VI]       & C\^{}m\^{}m7[I]        & \^{}BbvMm7[VII]     & CvMvM7[I]       \\
\midrule
\multicolumn{7}{l}{\textit{Neutral tonic} (\textsf{NN7})} \\
\midrule
2  & CNN7[I]         & CNN7/vvE[I]        & \^{}\^{}F\o\^{}m7[IV] & vvAnN7[VI]           & \^{}D\o vm7[II]     & \^{}Gn\^{}m7[V] \\
4  & CNN7[I]         & vvENN7/\^{}G[III]  & vvAnN7[VI]         & CNN7[I]                & \^{}\^{}F\o\^{}m7[IV] & vvBNm7[VII]  \\
\midrule
\multicolumn{7}{l}{\textit{Minor tonic} (\textsf{\^{}m\^{}m7})} \\
\midrule
6  & C\^{}m\^{}m7[I] & vD\^{}m\^{}m7/F[II] & vGvM\^{}m7[V]     & vvB\o\^{}m7[VII]       & D\#[III]            & vA\o 7[VI]      \\
8  & C\^{}m\^{}m7[I] & CvM\^{}m7/vE[I]    & vvF\#\o\^{}m7[IV]  & vA\^{}m\^{}m7[VI]      & vDvM\^{}m7[II]      & G\^{}mN7[V]     \\
\midrule
\multicolumn{7}{l}{\textit{Dominant tonic} (\textsf{vMvm7})} \\
\midrule
5  & CvMvm7[I]       & C\^{}m\^{}m7/D\#[I] & \^{}Fvm7[IV]      & vG\#maj7[VI]           & D\o 7[II]           & G\^{}m\^{}m7[V] \\
7  & CvMvm7[I]       & \^{}DvMvm7/F\#[II] & \^{}GvMvM7[V]      & B\^{}m\^{}m7[VII]      & E\^{}m\^{}m7[III]   & \^{}Avm7[VI]    \\
9  & CvMvm7[I]       & G\#vMvm7/C[VI]     & C\#vMvM7[II]       & F\^{}m\^{}m7[IV]       & Bb\^{}m\^{}m7[VII]  & D\#vm7[III]     \\
\bottomrule
\end{tabular}
\end{table}

% --------------------------------------------------------------------------------------------
\section{Discussion and Conclusion}
\label{sec:conclusion}

We address the discussion and conclusion in two sections: the first discusses the application and its Eigenspace, and the second the findings of the listening test.

\subsection{The Harmonic Eigenspace}

The geometric and topological models of harmony reviewed in Section~\ref{sec:literature_review} share a common dependency: their spatial structure is derived from music-theoretic axioms (fifth-relatedness in the Tonnetz, voice-leading parsimony in Tymoczko's chord spaces, scale membership in Balzano's group-theoretic model) that encode the properties of 12-TET as foundational assumptions. The Harmonic Eigenspace takes a different starting point. Its geometry is constructed from the psychoacoustic dissonance function alone, with no prior commitment to a tuning system, a scale, or a definition of consonance borrowed from tonal theory. The consequence is that 12-TET, 31-TET, and 53-TET chord types coexist in the same continuous volume and can be compared on the same perceptual basis.

This matters because the question of what a chord quality \emph{is} receives a specific answer in the Eigenspace formulation. The coordinate triple $(\alpha, \beta, \gamma)$ encodes quality as an invariant direction under uniform frequency scaling (transposition); the scale factor records the transposition interval. Two chords that share interval ratios but differ in root are the same point; two chords with different ratios are different points, regardless of what tuning system produced them. This is not a classification imposed from outside. It is a coordinate that falls out of the dissonance function, and it resolves a problem that existing spatial models do not address: how to locate and compare chord types across tuning systems that share no common scale structure.

The 100 local minima serve as the landmark system of the Eigenspace, partitioning it into basins that group chord types by their nearest just-intonation sonority. Within each basin, tempered chords from different tuning systems cluster around the same just-intonation minimum but at measurably different distances from it. The 12-TET dominant seventh and its nearest 53-TET variants, for example, occupy distinct positions within the same basin, and their dissonance values quantify a difference that Western theory acknowledges informally (the dominant seventh is ``not quite in tune'') but cannot measure. The Eigenspace makes these distinctions navigable and audible.

The harmonic exploration of new dissonant/consonant chord relations is open, as well as the definition of functional rules of chords in 53-TET. The dominant chord in 12-TET used by different cultural musical practices could be interpreted with different roles; in blues, the blue note implicitly bends towards a neutral quality \citep{cutting2018microtonal}, which causes the effect of making the chord \textit{solid} as a sonority, \textit{stronger}, or less dissonant; it could be perceived as more emphatic. If we find the closest stable zone, we find a just-intonation version that is very well-structured in harmonic terms, losing the 12-TET unstable quality. This observation is mostly perceptual; the only proposed test is to play with them in the web-app to explore the differences. 

Recent work by \citet{frank2026modeling} supports this reading. Applying variance decomposition to pleasantness ratings of 68 chords in 12-TET, they find that spectral interference is the primary axis of systematic between-listener variation in chord preferences, and that this effect strengthens with musical expertise. Their structural equation modelling also shows that preference variance increases with familiarity and decreases with interference: listeners agree on rough chords (they dislike them) but diverge on smooth, familiar ones. This aligns with our observation that neutral chords, which have no 12-TET counterpart, received dissonance ratings above their Eigenspace values, while consonant chords near 12-TET equivalents tracked theirs closely. Their finding that major-scale compatibility subsumes harmonicity as a predictor of pleasantness suggests that listeners respond to interval combinations they frequently encounter in music, rather than solely to spectral properties. The Eigenspace captures the spectral component without presupposing a scale system; the cultural component may account for the deviations.

\subsection{Listening Test}
\subsubsection{Part 1 - Chord Dissonance}

The category-level ordering largely held: Consonant chords were perceived as least dissonant (M = 3.37), followed by Intermediate (M = 4.89). However, two discrepancies are notable. Neutral chords were perceived as the most dissonant category (M = 6.40), substantially above their Eigenspace value (scaled = 5.28), while Diminished chords (M = 5.61) fell well below theirs (scaled = 7.21). Dominant chords showed a milder mismatch: perceived dissonance (M = 5.06) exceeded the Eigenspace value (scaled = 4.66). These deviations suggest that familiarity plays a role; listeners may penalise chord qualities that lack any approximation in 12-TET (such as neutral intervals) more than sensory roughness alone accounts for, while qualities closer to familiar tonal functions (such as dominant-type chords) may carry cultural expectations that inflate or deflate dissonance judgements independently of spectral roughness. For example, the $m7$ and $maj7$ are perceived with lower dissonance than their Eigenspace values. 

Inter-rater agreement was highest for chords closest to 12-TET equivalents (\textsf{m7}, SD = 1.57; \textsf{vMvM7}, SD = 1.74) and lowest for \textsf{vM\^{}m7} (Dominant; SD = 2.79) and \o 7 (Diminished; SD = 2.69). This pattern is consistent with the hypothesis that listeners share perceptual anchors for chord qualities they can map onto prior tonal experience, while chords occupying ambiguous territory between familiar categories produce greater individual variation.

Neutral qualities are in both Part 1 and Part 2 of the test perceived as dissonant, with the particularity that \textsf{smsm7} is the chord with the highest dissonance scores. The Eigenspace shows that \textsf{smsm7} is a higher-dissonance node compared to the counterpart \textsf{\^{}m\^{}m7}, which is the closest node to local-minimum 3. From the listening perception report, listeners perceive locality in both chords \textsf{\^{}m\^{}m7} $\rightarrow$ local-minimum 3 and \textsf{vMvM7} $\rightarrow$ local-minimum 1. As shown in Table~\ref{tab:lm_distances} and Figure~\ref{fig:chords_mapping_2}, both consonant chords sit virtually on top of their closest LM: \textsf{\^{}m\^{}m7} $\rightarrow$ LM-3 (dist $=$ 0.0015) and \textsf{vMvM7} $\rightarrow$ LM-1 (dist $=$ 0.0016). Listeners perceive locality here: these are the two most consonant chords. In contrast, \textsf{smsm7} $\rightarrow$ LM-10 (dist $=$ 0.0530) is much farther from its closest minimum than \textsf{\^{}m\^{}m7} is from LM-3, and LM-10 has a higher dissonance ($D = 12.95$) compared to LM-3 ($D = 11.60$) or LM-1 ($D = 11.73$). Overall, consonant chords sit at a distance $< 0.002$ from their closest LM, while neutral, less consonant, and diminished chords lie $0.03$--$0.07$ away, on the slopes, not at the bottom of the valleys.

Per-chord mean perceived dissonance ratings were compared against the dissonance values computed in the Eigenspace. Both correlation estimates are positive and their intervals exclude zero (Spearman $\rho = .481$, $p = .037$, approximate 95\% CI $[.01, .78]$; Pearson $r = .569$, $p = .011$, 95\% CI $[.15, .81]$; $n = 19$ chords), but the intervals are wide: nineteen chords support the direction of the association while leaving its magnitude undetermined, and we read the result as preliminary evidence. These were single planned comparisons, so no correction for multiple testing was applied. The Eigenspace category-level ordering (Consonant = 2.15, Intermediate = 4.54, Dominant = 4.66, Neutral = 5.28, Less Consonant = 5.52, Diminished = 7.21) broadly aligned with participants' perceived dissonance ordering, though Dominant chords showed a notable discrepancy, perceived as more dissonant than their Eigenspace values indicate. 
These deviations have more than one possible source: cultural familiarity, the timbre of the synthesised stimuli (sine and triangle waveforms), and the order of presentation. The study is correlational and cannot tell these sources apart. 

\subsubsection{Part 2 - Chord Progressions}
In Part~2, the overall mean ratings across the four dimensions were: coherence (M = 6.86), musicality (M = 6.43), novelty (M = 6.12), and dissonance (M = 5.60). These values indicate that listeners generally perceived the 53-TET progressions as coherent and musical, while also recognising them as novel.

Musicality as a concept encompasses creativity, cultural practice, rhythm, aesthetics, and performance, among other dimensions. The musicality ratings reported here do not attempt to capture the scope of musicality; they specifically measure the listener's perceptual tolerance for microtonal chord progressions. Musicality ratings varied substantially across progressions, ranging from M = 4.23 (Progression~4, Neutral tonic) to M = 8.00 (Progression~5, Dominant tonic). The three Major-tonic progressions (1, 3, 10) and the three Dominant-tonic progressions (5, 7, 9) received the highest musicality ratings, while the two Neutral-tonic progressions (2, 4) received the lowest. This ordering suggests that progressions built on tonic chords with closer approximations to 12-TET qualities (major and dominant) were perceived as more musical than those built on neutral chord types, which have no 12-TET counterpart.

The inter-dimension correlation analysis revealed consistent patterns across all four chord types. Coherence and musicality were positively correlated in every group (Major: $\rho = .784$; Dominant: $\rho = .691$; Minor: $\rho = .652$; Neutral: $\rho = .467$), indicating that listeners who found a progression coherent also tended to rate it as musical. Dissonance was negatively correlated with both coherence and musicality across all types, consistent with lower perceived musical quality at higher perceived dissonance. Novelty and musicality correlated positively for Major ($\rho = .395$) and Dominant ($\rho = .477$) types, and negligibly for Minor ($\rho = .140$), suggesting that for the more familiar-sounding tonic qualities, novelty was experienced as a positive musical attribute. The Neutral type was the exception: novelty and musicality were negatively (though not significantly) correlated ($\rho = -.247$), and novelty was positively correlated with dissonance ($\rho = .410$), suggesting that for progressions built on the most unfamiliar tonic quality, novelty and dissonance became perceptually entangled. The ratings are descriptive measures of acceptance/tolerance of first-encounter microtonal chord progressions, with no applicable reference, shaped by the listeners' cultural background, and we read them as exploratory. That listeners nonetheless differentiated degrees of novelty among uniformly unheard materials is itself part of the result.

\subsection{Limitations and Future Work}

The listening study has several limitations. The sample was small and uneven in gender composition; prior experience with microtonal music was recorded in the entry questionnaire but was not entered as a covariate, so the familiarity-based interpretations above remain tentative. The dissonance scale was anchored to two reference chords (maj6 as 1 and M+sm7 as 10); this practical choice constrains the ratings, which should not be read as independent of the chosen anchors. Finally, the reported correlations carry wide confidence intervals over 19 chords, and Part 2 is a reference-free, first-encounter design, so the perceptual results are exploratory and descriptive throughout.

The Eigenspace is computed here for harmonic timbres, where overtone frequencies are integer multiples of the fundamental. This is the overtone structure of strings and pipes, and the reason the LM align with just-intonation intervals. The dissonance function, however, is not restricted to this case. Instruments with inharmonic spectra (bells, metallophones, bars) produce entirely different dissonance topologies, and musical traditions built around such instruments use scales and intervals that correspond to those different surfaces. Computing the Eigenspace for inharmonic timbres would produce a different set of LM, different basins of reference, and a different chord vocabulary. The mathematical structure of the Eigenspace applies unchanged to any spectrum: the coordinates, the dissonance field, and the extraction of minima take the partials as input, and the harmonic content revealed is determined entirely by the timbre. This observation points toward a broader programme: a family of Eigenspaces, each generated by a different spectral profile, mapping the perceptual harmonic possibilities specific to each timbral world. Recent work by \citet{etxebarria2025dissonance} demonstrate this concretely for the one-dimensional case: dissonance curves computed from the spectra of ancient idiophones and aerophones recover the scales of Thai classical music and Basque traditional practice, confirming that the relationship between timbre and tuning holds across culturally independent musical traditions. Extending this analysis to the four-dimensional tetrad case would produce a family of Eigenspaces, each mapping the harmonic possibilities specific to a given timbral world.

Such a programme would ground comparative musicology in a common analytical framework, connecting tuning systems that appear unrelated when viewed through the lens of Western tonal theory but that may share structural features when viewed through the lens of their respective dissonance topologies. Within the present Eigenspace, the chord vocabulary itself will grow: applying the ten-fold quality gradation to the fifth, as it already applies to the third and the seventh, completes the combinatorial family and is planned for the application. The coordinate system also lends itself to a spatial grounding of chord quality for machine-learning models of microtonal harmony; a companion study (anonymous) pursues this direction.

\section*{Acknowledgement}
Hidden during the review process.
% This project has received funding from the European Union's Horizon Europe research and innovation programme under the Marie Sk\l{}odowska-Curie Actions (MSCA) Postdoctoral Fellowship, grant agreement No.\ 101203318. The authors thank the participants who volunteered for the listening study, and S\'onar+D for exhibiting the project and presenting the music composed using the proposed system.

% An unnumbered section, e.g.\ \verb"\section*{Acknowledgements}", may be used for thanks, etc.\ if required and included \emph{in the non-anonymous version} before any Notes or References.

\section*{Disclosure statement}
The authors have no financial or commercial relationships that could be construed as a potential conflict of interest.

\noindent \textbf{Use of Generative AI:} The authors used Anthropic Claude Opus 4.6 via the claude.ai platform for proofreading, including grammar correction, typo fixes, and stylistic consistency of the manuscript text. No content was generated by the model. The intellectual content, research design, analysis, and interpretation remain entirely the authors' own.

\section*{Funding}
Hidden during the review process.
% This work was supported by the European Union's Horizon Europe research and innovation programme under the Marie Sk\l{}odowska-Curie Actions grant agreement No.\ 101203318 (ANIMA: Artificial INtelligence-based Interactive Microtonal Compositional Assistant).

% \section*{Notes on contributor(s)}

% An unnumbered section, e.g.\ \verb"\section*{Notes on contributors}", may be included \emph{in the non-anonymous version} if required. A photograph may be added if requested.

% \section*{Nomenclature/Notation}

% An unnumbered section, e.g.\ \verb"\section*{Nomenclature}" (or \verb"\section*{Notation}"), may be included if required, before any Notes or References.

% \section*{Notes}

% An unnumbered `Notes' section may be included before the References (if using the \verb"endnotes" package, use the command \verb"\theendnotes" where the notes are to appear, instead of creating a \verb"\section*").

\bibliographystyle{apalike}
\bibliography{references}

\appendix
\section{Application usage details}
\label{app:usage}

This appendix collects the operational details of the interface referenced in Section~\ref{sec:application}.

\begin{figure}[H]
    \centering
    \includegraphics[width=0.4\columnwidth]{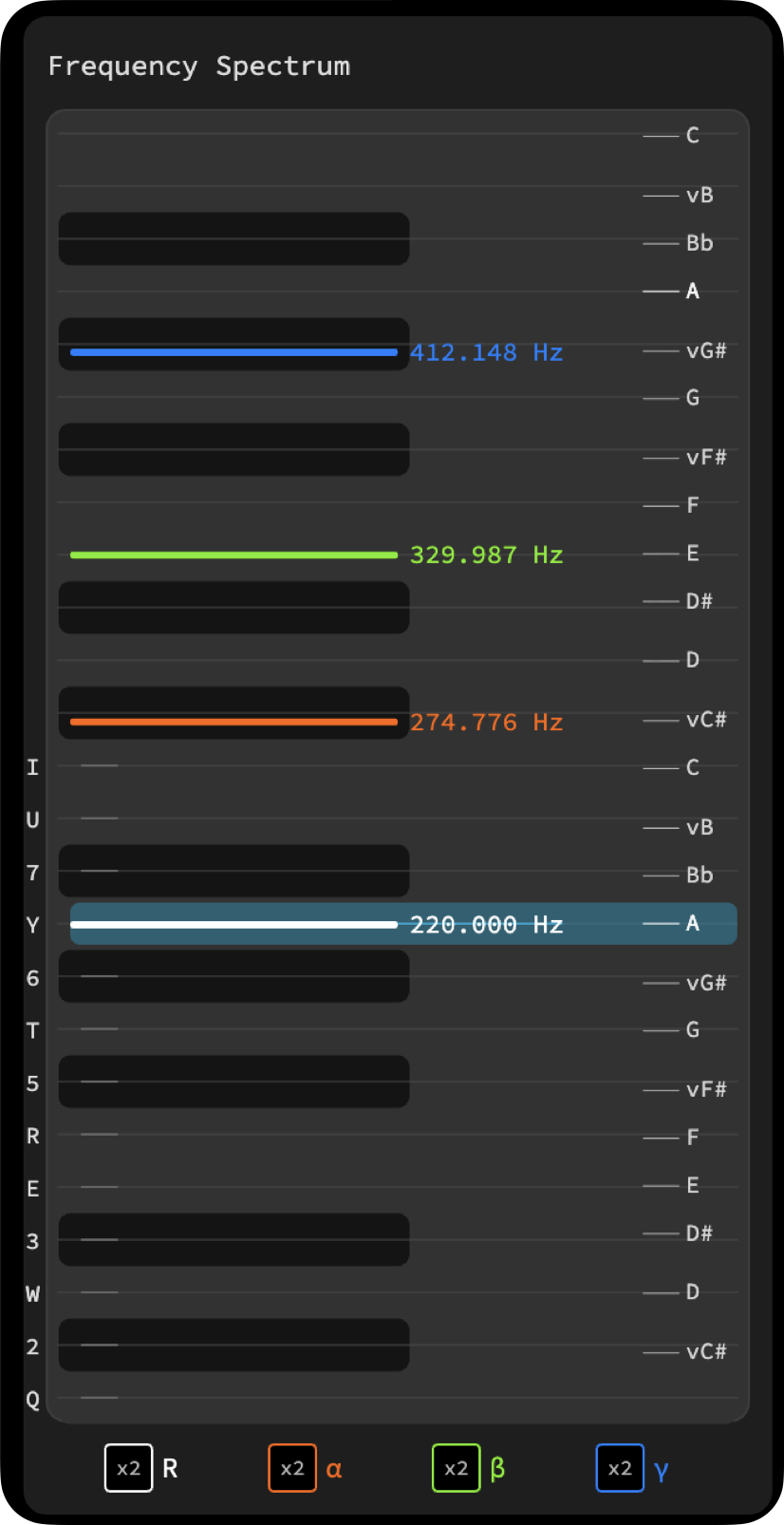}
    \caption{Frequency spectrum visualisation showing the deviation from standard 12-TET positions on a piano-roll reference.}
    \label{fig:piano_visualization}
\end{figure}

\paragraph*{Frequency Spectrum Visualisation:}
When a node is clicked, the tetrad is displayed on a piano-roll reference showing the deviation from standard 12-TET positions, as shown in Figure~\ref{fig:piano_visualization}. Each axis is colour-coded: root = white, $\alpha$ = red, $\beta$ = green, $\gamma$ = blue. Each axis has a `\textit{\textbf{$\times2$}}' button to arrange different voicing formatting, moving the corresponding note one octave higher. The root of the related node sonification can be changed by clicking the piano location of the new root or by pressing the computer keyboard from the key `Q' to `I'. 

\begin{figure}[H]
    \centering
    \includegraphics[width=0.5\columnwidth]{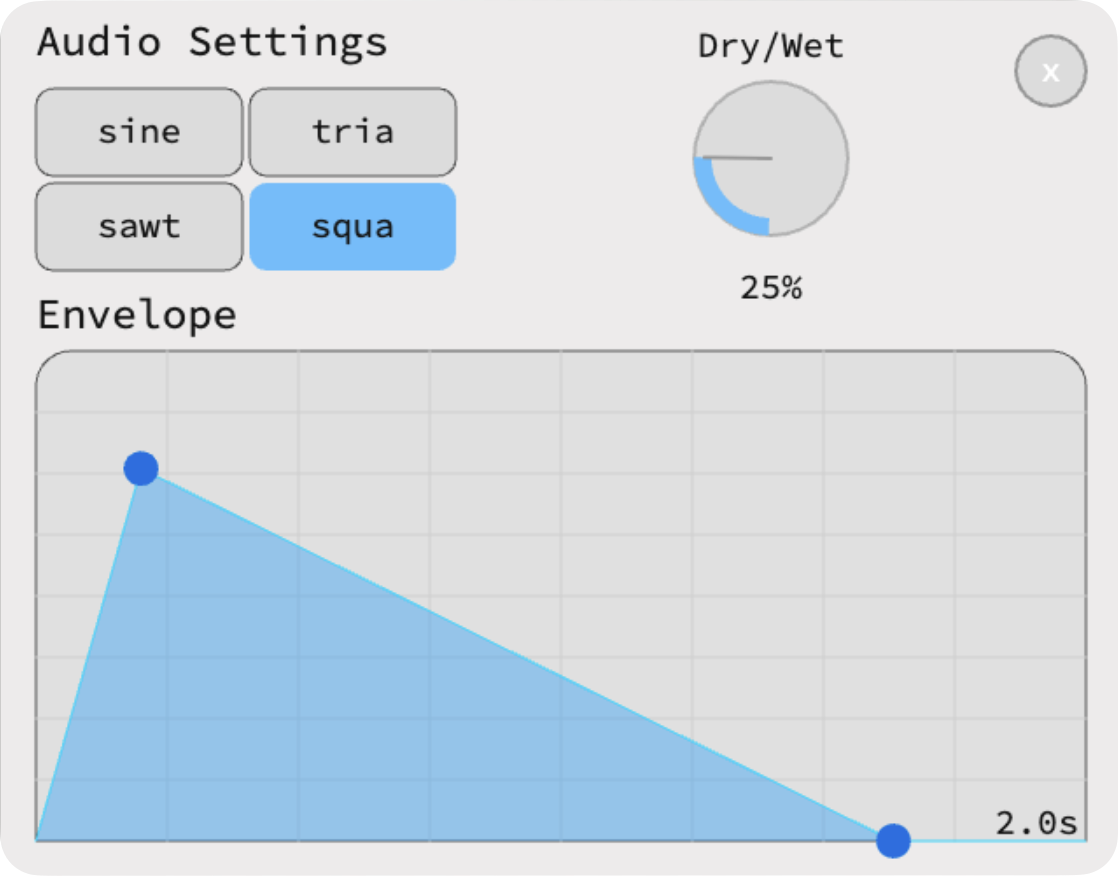}
    \caption{Audio engine settings. Waveform selection (sine, triangle, sawtooth, square), reverb control, and an envelope graph for attack, sustain, and release timing.}
    \label{fig:audio_settings}
\end{figure}

\begin{figure}[H]
    \centering
    \includegraphics[width=0.5\columnwidth]{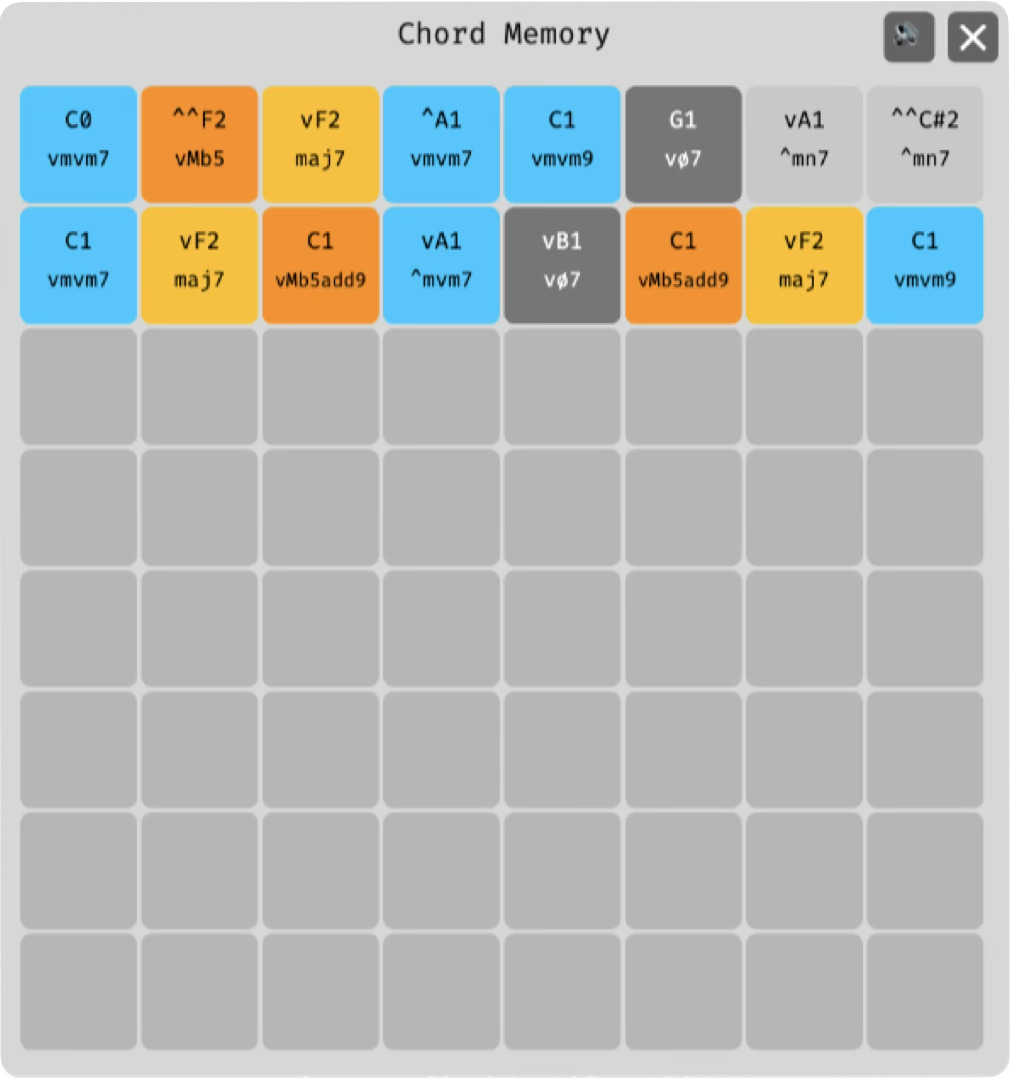}
    \caption{The Chord Memory grid is used to store data from nodes of any type. It serves as a chord progression or sequence to play with an external audio workstation.}
    \label{fig:chord_grid}
\end{figure}

\paragraph*{Audio Engine:}
The audio engine is programmed using the \textit{Web Audio API}\footnote{\url{https://developer.mozilla.org/en-US/docs/Web/API/Web_Audio_API}}, which allows generating waveforms such as sine, triangle, sawtooth, and square (see Figure~\ref{fig:audio_settings}). It has integrated reverb with a percentage knob and an envelope graph to control the attack, decay, and release timing within a 2-second range. The interface is programmed using the open-source library p5.js\footnote{\url{https://p5js.org}}, which is the web version of Processing\footnote {\url{https://processing.org}}. The chords can be stored in a \textit{Chord Memory} grid (see Figure~\ref{fig:chord_grid}) to form sequences or chord progressions, enabling interactive play with the web application as a MIDI controller in any DAW. The grid can be linked to the MIDI controller \textit{Launchpad Pro}\footnote{\url{https://novationmusic.com/launchpad}}. 

% \noindent\textbf{Appendix A. This is the title of the first appendix}\medskip

% \noindent\textbf{Appendix B. This is the title of the second appendix}\medskip

%\processdelayedfloats %%% See above for an explanation of why this command might be needed here.

% \appendix

\end{document}